\documentclass[12pt]{article}
\usepackage{epsfig,a4}
 \advance\oddsidemargin by -1in
 \advance\evensidemargin
 by -1in
 \marginparsep
 0.4cm
 \advance\topmargin by
 -0.0in
 \makeindex

 \newcommand\la{\langle}
 \newcommand\ra{\rangle}
 \newcommand\beq{\begin{equation}}
 
 \newcommand\eeq{\end{equation}}
 \newcommand\beqn{\begin{eqnarray}}
 \newcommand\eeqn{\end{eqnarray}}
 \newcommand\GeV{{\rm GeV}}

\def\BA{\begin{eqnarray}}
\def\BE{\begin{equation}}
\def\BF{\begin{figure}[htb]}
\def\BT{\begin{table}[htb]}
\def\EA{\end{eqnarray}}
\def\EE{\end{equation}}
\def\EF{\end{figure}}
\def\ET{\end{table}}

\def\la{\langle}
\def\ra{\rangle}
\def\mb{\,\mbox{mb}}
\def\fm{\,\mbox{fm}}
\def\GeV{\,\mbox{GeV}}

\def\Pom{{\bf I\!P}}

\def\lsim{\mathrel{\rlap{\lower4pt\hbox{\hskip1pt$\sim$}}
    \raise1pt\hbox{$<$}}}         %less than or approx. symbol
\def\gsim{\mathrel{\rlap{\lower4pt\hbox{\hskip1pt$\sim$}}
    \raise1pt\hbox{$>$}}}         %greater than or approx. symbol

\begin{document}
\vspace*{3cm}

\date{today}

\begin{center}
{\LARGE \bf
%%%%%%%%%%%%%%%%%%%%%%%%%%%%%%%%%%%%%%%%%
Gluon Shadowing in DIS off Nuclei
%%%%%%%%%%%%%%%%%%%%%%%%%%%%%%%%%%%%%%%%%
}
\end{center}

\begin{center}

\vspace{0.5cm}
 {\large B.Z.~Kopeliovich$^{1,2}$, J.~Nemchik$^{3,4}$,
I.K.~Potashnikova$^{1,2}$ and
Ivan~Schmidt$^{1}$}
 \\[1cm]
 {$^{1}$\sl Departamento de F\'{\i}sica y Centro de Estudios
Subat\'omicos,\\ Universidad T\'ecnica Federico Santa Mar\'{\i}a,
Valpara\'{\i}so, Chile} \\
 {$^{2}$\sl Joint Intitute for Nuclear Research, Dubna, Russia} \\
 {$^{3}$\sl
Institute of Experimental Physics SAS, Watsonova 47,
04001 Kosice, Slovakia} \\
{$^{4}$\sl Czech Technical University,
FNSPE, Brehova 7,
11519 Praque, Czech Republic}

\end{center}

\vspace{1cm}

\date{\today}

\vspace{1cm}
%----------------------------------------------------
\begin{abstract}

Within a light-cone quantum-chromodynamics dipole formalism based on
the Green function technique, we study nuclear shadowing in
deep-inelastic scattering at small Bjorken $x_{Bj}\lsim 0.01$. Such
a formalism incorporates naturally color transparency and coherence
length effects. Calculations of the nuclear shadowing for the $\bar
qq$ Fock component of the photon are based on an exact numerical
solution of the evolution equation for the Green function, using a
realistic form of the dipole cross section and nuclear density
function. Such an exact numerical solution is unavoidable for
$x_{Bj}\gsim 10^{-4}$, when  a variation of the transverse size of
the $\bar qq$ Fock component must be taken into account. The eikonal
approximation, used so far in most other models, can be applied only
at high energies, when $x_{Bj}\lsim 10^{-4}$ and the transverse size
of the $\bar qq$ Fock component is "frozen" during propagation
through the nuclear matter. At $x_{Bj}\le 0.01$ we find quite a
large contribution of gluon suppression to nuclear shadowing, as a
shadowing correction for the higher Fock states containing gluons.
Numerical results for nuclear shadowing are compared with the
available data from the E665 and NMC collaborations. 
Nuclear shadowing is also predicted at very small $x_{Bj}$ corresponding
to LHC kinematical range.  
Finally the model predictions are compared and discussed 
with the results obtained from other
models.

\end{abstract}
%----------------------------------------------------
%\doublespace
\newpage

%\pacs{13.85.Lg, 13.60.Le}

%\maketitle

%%%%%%%%%%%%%%%%%%%%%%%%%%%%%%%%%
\section{Introduction}
\label{intro}
%%%%%%%%%%%%%%%%%%%%%%%%%%%%%%%%%

Nuclear shadowing in deep-inelastic scattering (DIS) off nuclei is
usually studied via nuclear structure functions. In the shadowing
region of small Bjorken $x_{Bj}\lsim 0.01$ the structure function
$F_2$ per nucleon turns out to be smaller in nuclei than in a free
nucleon (see the review \cite{arneodo-94}, for example). This
affects then the corresponding study of nuclear effects, mainly in
connection with the interpretation of the results coming from
hadron-nucleus and heavy ion experiments.

Nuclear shadowing, intensively investigated during the last two
decades, can be treated differently depending on the reference
frame. In the infinite momentum frame of the nucleus it can be
interpreted as a result of parton fusion
\cite{k-73,glr-83,mq-86,q-87}, leading to a reduction of the parton
density at low Bjorken $x_{Bj}$. In the rest frame of the nucleus,
however, this phenomenon looks like nuclear shadowing of the
hadronic fluctuations of the virtual photon, and occurs due to their
multiple scattering inside the target
\cite{bsyp-78,fs-88,bl-90,bsy-04,nz-91,mt-93,npz-95,prw-95,kp-96,kp-97,pw-00,krt-00,n-03}.
Although these two physical interpretations are complementary, the
one based on the rest frame of the nucleus is more intuitive and
straightforward.

The dynamics of nuclear shadowing in DIS is controlled by the effect
of quantum coherence, which results from the destructive
interference of amplitudes for which the interaction takes place on
different bound nucleons. Taking into account the $|\bar qq\ra$ Fock
component of the photon, quantum coherence can be characterized by
the lifetime of the $\bar qq$ fluctuation, which in turn can be
estimated by relying on the uncertainty principle and Lorentz time
dilation as,
%
%%%%%%%%%%%%%%%%%%%%%%%%%%%%%%%%%%%
 \beq
t_c = \frac{2\,\nu}{Q^2 + M_{\bar qq}^2}\ ,
%----------
\label{10}
%----------
 \eeq
%%%%%%%%%%%%%%%%%%%%%%%%%%%%%%%%%%%
%
where $\nu$ is the photon energy, $Q^2$ is photon virtuality and
$M_{\bar qq}$ is the effective mass of the $\bar qq$ pair. This is
usually called coherence time, but we also will use the term
coherence length (CL), since light-cone kinematics is assumed,
$l_c=t_c$. The CL is related to the longitudinal momentum transfer
by $q_c=1/l_c$. Notice that for higher Fock states containing gluons
$|\bar qqG\ra$, $|\bar qq2G\ra$, ... , the corresponding effective
masses are larger than $M_{\bar qq}$. Consequently, these
fluctuations have a shorter coherence time than the lowest $|\bar
qq\ra$ state. The effect of CL is naturally incorporated in the
Green function formalism, which has been already applied to DIS,
Drell-Yan pair production \cite{krt-98,krt-00,n-03}, and vector
meson production \cite{knst-01,n-02} (see also the next Section).

In the present paper nuclear shadowing in DIS will be treated using
again the Green function approach. Such a quantum mechanical
treatment requires to solve the evolution equation for the Green
function. Usually, for simplicity this equation is set up in a such
way as to obtain the Green function in an analytical form (see
\cite{krt-98,krt-00}, for example), which requires, however, to
implement several approximations into a rigorous quantum-mechanical
approach, like a constant nuclear density function (\ref{270}) and a
specific quadratic form (\ref{260}) of the dipole cross section. The
solution obtained in a such way is the harmonic oscillator Green
function \cite{kz-91} (see also Eq.~(\ref{142})), usually used for
calculation of nuclear shadowing \cite{krt-98,krt-00,r-00}. Then the
question about the accuracy of the predictions for nuclear shadowing
using such approximations naturally arises.

In the process of searching for the corresponding answer, in 2003
the evolution equation for the Green function was solved numerically
for the first time in ref.~\cite{n-03}. This allowed to exclude any
additional assumptions and avoid supplementary approximations, which
caused theoretical uncertainties. The corresponding predictions for
nuclear shadowing in DIS at small $x_{Bj}$, based on the exact
numerical solution of the evolution equation for the Green function
\cite{n-03}, showed quite a large difference in comparison with
approximate calculations \cite{krt-98,krt-00} obtained within the
harmonic oscillator Green function approach, in the kinematic region
when $l_c \lsim R_A$ ($R_A$ is the nuclear radius). However, no
comparison with data was performed using this path integral
technique based on an exact numerical solution of the
two-dimensional Schr\" odinger equation for the Green function. This
is one of the main goals of the present paper. Such a comparison
with data provides a better baseline for future studies of the QCD
dynamics, not only in DIS off nuclei but also in further processes
occurring in lepton (proton)-nucleus interactions and in heavy-ion
collisions.

The calculations of nuclear shadowing in DIS off nuclei presented so
far within the light-cone (LC) Green function approach
\cite{krt-98,krt-00,n-03} were performed assuming only $\bar qq$
fluctuations of the photon, and neglecting higher Fock components
containing gluons and sea quarks. The effects of higher Fock states
are included in the energy dependence of the dipole cross section,
$\sigma_{\bar qq}(\vec{r},s)$\footnote{Here $\vec r$ represents the
transverse separation of the $\bar qq$ photon fluctuation and $s$ is
the center of mass energy squared.}. However, as soon as nuclear
effects are considered, these Fock states $|\bar qqG\ra$, $|\bar
qq2G\ra$ ..., lead to gluon shadowing (GS), which for simplicity has
been neglected so far when the model predictions were compared with
experimental data. The contribution of the gluon suppression to
nuclear shadowing represents a shadowing correction for the
multigluon higher Fock states. It was shown in ref.~ \cite{kst2}
that GS becomes effective at small $x_{Bj}\lsim 0.01$. The present
available experimental data cover the shadowing region $\sim
0.0001\lsim x_{Bj}\lsim 0.01$, and therefore the contribution of GS
to the overall nuclear shadowing should be included. This is a
further goal of the present paper.

Different (but equivalent) descriptions of GS are known, depending
on the reference frame. In the infinite momentum frame of the
nucleus it looks like fusion of gluons, which overlap in the
longitudinal direction at small $x_{Bj}$, leading to a reduction of
the gluon density. In the rest frame of the nucleus the same
phenomenon looks as a specific part of Gribov's inelastic
corrections \cite{gribov}. The lowest order inelastic correction
related to diffractive dissociation $\gamma^*\,N \to X\,N$ \cite{kk}
contains PPR and PPP contributions (in terms of the triple-Regge
phenomenology, see \cite{kklp}). The former is related to quark
shadowing, while the latter, the triple-Pomeron term, corresponds to
gluon shadowing. Indeed, only diffractive gluon radiation can
provide the $M_X$ dependence $d\sigma_{dd}/dM^2_X \propto 1/M_X^2$
of the diffractive dissociation cross section. In terms of the
light-cone QCD approach the same process is related to the inclusion
of higher Fock components, $|\bar qq\,nG\ra$, containing gluons
\cite{al}. Such fluctuations might be quite heavy compared to the
simplest $|\bar qq\ra$ fluctuation, and therefore have a shorter
lifetime (see Eq.~(\ref{10})), and need higher energies to be
relevant.

Calculations of the GS contribution to nuclear suppression have been
already performed within the light-cone QCD approach, for both
coherent and incoherent production of vector mesons
\cite{knst-01,n-02}, and also for production of Drell-Yan pairs
\cite{krt-00}. They showed (except for the specific case of
incoherent production of vector mesons)  that GS is a non-negligible
effect, especially for heavy nuclear targets at small and medium
values of photon virtualities $Q^2\lsim$ a few GeV$^2$ and at large
photon energies $\nu$. This is another reason to include the effect
of GS  for the calculation of nuclear shadowing, especially for
making more realistic comparison of the predictions with
experimental data.

Notice also that by investigating shadowing in the region of small
$x_{Bj}\lsim 0.01$ we can safely omit the nuclear antishadowing
effect assumed to be beyond the shadowing dynamics
\cite{bl-90,bsy-04}.

The paper is organized as follows. In the next Section~\ref{lc} we
present a short description of the light-cone dipole phenomenology
for nuclear shadowing in DIS, together with the Green function
formalism. In Section~\ref{glue-shadow} we discuss how gluon
shadowing modifies the total photoabsorption cross section on a
nucleus. In Section~\ref{results} numerical results are presented
and compared with experimental data, and also with the results from
other models, in a broad range of $x_{Bj}$. Finally, in
Section~\ref{conclusions} we summarize our main results and discuss
the possibility of future experimental evidence of the GS
contribution to the overall nuclear shadowing in DIS at small values
of $x_{Bj}$.

%
%%%%%%%%%%%%%%%%%%%%%%%%%%%%%%%%%%%%%%%%%%%%%%%%%%%%%%%%%%%%%%%%%%%%%%%%%%%
\section{Light-cone dipole approach to nuclear
shadowing}
\label{lc}
%%%%%%%%%%%%%%%%%%%%%%%%%%%%%%%%%%%%%%%%%%%%%%%%%%%%%%%%%%%%%%%%%%%%%%%%%%%
%

%The main goal
%One clear advantage of the LC dipole approach to nuclear shadowing
%is the possibility to include
%%the nuclear
%nucleon form factors in all multiple scattering terms.
In the rest frame of the nucleus the nuclear shadowing in the total
virtual photoabsorption cross section
$\sigma_{tot}^{\gamma^*A}(x_{Bj},Q^2)$ (or in the structure function
$F_2^A(x_{Bj},Q^2)$) can be decomposed over different Fock
components of the virtual photon. Then the total photoabsorption
cross section on a nucleus can be formally represented in the form
%
%%%%%%%%%%%%%%%%%%%%%%%%%%%%%%%%%%%%%%%%%%%%%%%%%%%%%%%%%%%
 \BE
\sigma_{tot}^{\gamma^*A}(x_{Bj},Q^2) =
A~\sigma_{tot}^{\gamma^*N}(x_{Bj},Q^2) -
\Delta\sigma_{tot}(x_{Bj},Q^2)\, ,
%----------
\label{110}
%----------
 \EE
%%%%%%%%%%%%%%%%%%%%%%%%%%%%%%%%%%%%%%%%%%%%%%%%%%%%%%%%%%%
%
where
%
%%%%%%%%%%%%%%%%%%%%%%%%%%%%%%%%%%%%%%%%%%%%%%%%%%%%%%%%%%%
 \BE
\Delta\sigma_{tot}(x_{Bj},Q^2) =
\Delta\sigma_{tot}(\bar qq) +
\Delta\sigma_{tot}(\bar qqG) +
\Delta\sigma_{tot}(\bar qq2G) + \cdot
\, .
%----------
\label{112}
%----------
 \EE
%%%%%%%%%%%%%%%%%%%%%%%%%%%%%%%%%%%%%%%%%%%%%%%%%%%%%%%%%%%
%
Here the Bjorken variable $x_{Bj}$ is given by
%
%%%%%%%%%%%%%%%%%%%%%%%%%%%%%%%%%%%%%%%%
\beq
x_{Bj} = \frac{Q^2}{2\,m_N\,\nu} \approx
\frac{Q^2}{Q^2 + s}\, ,
%----------
\label{115}
%----------
\eeq
%%%%%%%%%%%%%%%%%%%%%%%%%%%%%%%%%%%%%%%%
%
where $s$ is the $\gamma^*$-nucleon center of mass (c.m.) energy
squared, $m_N$ is mass of the nucleon, and
$\sigma_{tot}^{\gamma^*N}(x_{Bj},Q^2)$ in (\ref{110}) is total
photoabsorption cross section on a nucleon
%
%%%%%%%%%%%%%%%%%%%%%%%%%%%%%%%%%%%%%%%%%%%%%%%%%%%%%%%%%%%
 \BE
\sigma_{tot}^{\gamma^*N}(x_{Bj},Q^2) =
\int d^2 r \int_{0}^{1} d\alpha\,\Bigl
| \Psi_{\bar qq}(\vec{r},\alpha,Q^2)\,\Bigr |^2
~\sigma_{\bar qq}(\vec{r},s)\, .
%----------
\label{120}
%----------
 \EE
%%%%%%%%%%%%%%%%%%%%%%%%%%%%%%%%%%%%%%%%%%%%%%%%%%%%%%%%%%%
%
In this last expression $\sigma_{\bar qq}({\vec{r}},s)$ is the
dipole cross section, which depends on the $\bar qq$ transverse
separation $\vec{r}$ and the c.m. energy squared $s$, and
$\Psi_{\bar qq}({\vec{r}},\alpha,Q^2)$ is the LC wave function of
the $\bar qq$ Fock component of the photon, which depends also on
the photon virtuality $Q^2$ and the relative share $\alpha$ of the
photon momentum carried by the quark. Notice that $x_{Bj}$ is
related to the c.m. energy squared $s$ via Eq.~(\ref{115}).
Consequently, hereafter we will write the energy dependence of
variables in subsequent formulas also via an $x_{Bj}$-dependence
whenever convenient.

The total photoabsorption cross section on a nucleon target
(\ref{120}) contains two ingredients.  The first ingredient is given
by the dipole cross section $\sigma_{\bar qq}(\vec r,s)$,
representing the interaction of a $\bar qq$ dipole of transverse
separation $\vec r$ with a nucleon \cite{zkl}. It is a flavor
independent universal function of $\vec{r}$ and energy, and allows
to describe various high energy processes in an uniform way. It is
also known to vanish quadratically $\sigma_{\bar qq}(r,s)\propto
r^2$ as $r\rightarrow 0$, due to color screening (property of color
transparency \cite{zkl,bbgg,bm-88}), and cannot be predicted
reliably because of poorly known higher order perturbative QCD
(pQCD) corrections and nonperturbative effects. However, it can be
extracted from experimental data on DIS and structure functions
using reasonable parametrizations, and in this case pQCD corrections
and nonperturbative effects are naturally included in $\sigma_{\bar
qq}(r,s)$.

There are two popular parameterizations of $\sigma_{\bar qq}(\vec
r,s)$: GBW presented in \cite{gbw}, and KST proposed in \cite{kst2}.
Detailed discussions and comparison of these two parametrizations
can be found in refs.~\cite{r-00,knst-01,n-03}. Whereas the GBW
parametrization cannot be applied in the nonperturbative region of
$Q^2$, the KST parametrization gives a good description of the
transition down to the limit of real photoproduction, $Q^2=0$.
Because we will study the shadowing region of small $x_{Bj}\lsim
0.01$, where available experimental data from the E665 and NMC
collaborations cover small and moderate values of $Q^2\lsim 2\div
3$\,GeV$^2$, we will prefer the latter parametrization.

The KST parametrization \cite{kst2} has the following form, which
contains an explicit dependence on energy,
%
%%%%%%%%%%%%%%%%%%%%%%%%%%%%%%%%
\beq
\sigma_{\bar qq}(r,s) = \sigma_0(s)\,\left [1 -
exp\left ( - \frac{r^2}{R_0^2(s)}\right )\right ]\, .
%-----------
\label{kst-1}
%-----------
\eeq
%%%%%%%%%%%%%%%%%%%%%%%%%%%%%%%%
%
The explicit energy dependence in the parameter $\sigma_0(s)$ is
introduced in a such way that it guarantees that the correct
hadronic cross sections is reproduced,
%
%%%%%%%%%%%%%%%%%%%%%%%%%%%%%%%%%
\beq
\sigma_0(s) = \sigma_{tot}^{\pi\,p}(s)\,\left
(1 + \frac{3\,R_0^2(s)}{8\,\la r_{ch}^2\ra_{\pi}}\right )\, ,
%------------
\label{kst-2}
%------------
\eeq
%%%%%%%%%%%%%%%%%%%%%%%%%%%%%%%%%
%
where $\sigma_{tot}^{\pi\,p}(s) = 23.6\,(s/s_0)^{0.079} +
1.432\,(s/s_0)^{-0.45}\mb$, which contains the Pomeron and Reggeon
parts of the $\pi p$ total cross section \cite{rpp-96}, and $R_0(s)
= 0.88\,(s/s_0)^{-\lambda/2}\fm$, with $\lambda = 0.28$ and where
$s_0 = 1000\GeV^2$ is the energy-dependent radius. In
Eq.~(\ref{kst-2}) $\la r_{ch}^2\ra_{\pi} = 0.44\fm^2$ is the mean
pion charge radius squared. The form of Eq.~(\ref{kst-1})
successfully describes the data for DIS at small $x_{Bj}$ only up to
$Q^2\approx 10\GeV^2$. Nevertheless, this interval of $Q^2$ is
sufficient for the purpose of the present paper, which is focused on
the study of nuclear shadowing at small $x_{Bj}\lsim 0.01$ in the
kinematic range $Q^2\lsim 4\,\GeV^2$ covered by available E665 and
NMC data. \\
However, as we will present the predictions for nuclear shadowing
at very small $x_{Bj}$ down to $10^{-7}$ accesible
by the prepared experiments at LHC and at larger values
of $Q^2\gsim 10\,\GeV^2$, we will use also the second GBW parametrization
\cite{gbw} of the dipole cross section.

The second ingredient of $\sigma_{tot}^{\gamma^*N}(x_{Bj},Q^2)$ in
(\ref{120}) is the perturbative distribution amplitude (``wave
function'') of the $\bar qq$ Fock component of the photon. For
transversely (T) and longitudinally (L) polarized photons it has the
form \cite{lc,bks-71,nz-91}:
%
%%%%%%%%%%%%%%%%%%%%%%%%%%%%%%%%%%%%%%%%%%%%%%
 \BE
\Psi_{\bar qq}^{T,L}({\vec{r}},\alpha,Q^2) =
\frac{\sqrt{N_{C}\,\alpha_{em}}}{2\,\pi}\,\,
Z_{q}\,\bar{\chi}\,\hat{O}^{T,L}\,\chi\,
K_{0}(\epsilon\,r)
%----------
\label{122}
%----------
 \EE
%%%%%%%%%%%%%%%%%%%%%%%%%%%%%%%%%%%%%%%%%%%%%%
%
where $\chi$ and $\bar{\chi}$ are the spinors of the quark and
antiquark respectively, $Z_{q}$ is the quark charge, $N_{C} = 3$ is
the number of colors, and $K_{0}(\epsilon r)$ is a modified Bessel
function with
%
%%%%%%%%%%%%%%%%%%%%%%%%%%%%%%%%%%%%%%%%%%%%
 \BE
\epsilon^{2} =
\alpha\,(1-\alpha)\,Q^{2} + m_{q}^{2}\ ,
%----------
\label{123}
%----------
 \EE
%%%%%%%%%%%%%%%%%%%%%%%%%%%%%%%%%%%%%%%%%%%%
%
where $m_{q}$ is the quark mass. The operators $\widehat{O}^{T,L}$
read,
%
%%%%%%%%%%%%%%%%%%%%%%%%%%%%%%%%%%%%%%%%%%%%%%%%%%%%%%%%%%%%
 \BE
\widehat{O}^{T} = m_{q}\,\,\vec{\sigma}\cdot\vec{e} +
i\,(1-2\alpha)\,(\vec{\sigma}\cdot\vec{n})\,
(\vec{e}\cdot\vec{\nabla}_r) + (\vec{\sigma}\times
\vec{e})\cdot\vec{\nabla}_r\ ,
%-----------
 \label{124}
%-----------
 \EE
%%%%%%%%%%%%%%%%%%%%%%%%%%%%%%%%%%%%%%%%%%%%%%%%%%%%%%%%%%%%
%
%
%%%%%%%%%%%%%%%%%%%%%%%%%%%%%%%%%%%%%%%%%%%%%%%%%%%%%%%%%%%%
 \BE
\widehat{O}^{L} =
2\,Q\,\alpha (1 - \alpha)\,(\vec{\sigma}\cdot\vec{n})\ .
%----------
\label{125}
%----------
 \EE
%%%%%%%%%%%%%%%%%%%%%%%%%%%%%%%%%%%%%%%%%%%%%%%%%%%%%%%%%%%%
%
 Here $\vec\nabla_r$ acts on the transverse coordinate $\vec r$,
$\vec{e}$ is the polarization vector of the photon, $\vec{n}$ is a
unit vector parallel to the photon momentum, and $\vec{\sigma}$ is
the three vector of the Pauli spin-matrices.

The distribution amplitude Eq.~(\ref{122}) controls the transverse
$\bar qq$ separation with the mean value
%
%%%%%%%%%%%%%%%%%%%%%%%%%%%%%%%%%%%%%%%%%%%%%%%%%%%%%%%%
 \BE
\la r\ra \sim \frac{1}{\epsilon} =
\frac{1}{\sqrt{Q^{2}\,\alpha\,(1-\alpha) + m_{q}^{2}}}\,.
\label{130}
 \EE
%%%%%%%%%%%%%%%%%%%%%%%%%%%%%%%%%%%%%%%%%%%%%%%%%%%%%%%%
%

 For very asymmetric $\bar qq$ pairs with $\alpha$ or $(1-\alpha) \lsim
m_q^2/Q^2$ the mean transverse separation $\la r\ra \sim 1/m_q$
becomes huge, since one must use current quark masses within pQCD. A
popular recipe to fix this problem is to introduce an effective
quark mass $m_{eff}\sim \Lambda_{QCD}$, which represents the
nonperturbative interaction effects between the $q$ and $\bar q$. It
is more consistent and straightforward, however, to introduce this
interaction explicitly through a phenomenology based on the
light-cone Green function approach, and which has been developed in
\cite{kst2}.

The Green function $G_{\bar qq}(\vec{r_2},z_2;\vec{r_1},z_1)$
describes the propagation of an interacting $\bar qq$ pair between
points with longitudinal coordinates $z_1$ and $z_2$ and with
initial and final separations $\vec{r_1}$ and $\vec{r_2}$. This
Green function satisfies the two-dimensional Schr\"odinger equation,
%
%%%%%%%%%%%%%%%%%%%%%%%%%%%%%%%%%%%%%%%%%%%%%%%%%%%%%%%%%%%%%%%%%%%%%%%%
 \BE
i\frac{d}{dz_2}\,G_{\bar qq}(\vec{r_2},z_2;\vec{r_1},z_1)=
\left[\frac{\epsilon^{2} - \Delta_{r_{2}}}{2\,\nu\,\alpha\,(1-\alpha)}
+V_{\bar qq}(z_2,\vec{r_2},\alpha)\right]
G_{\bar qq}(\vec{r_2},z_2;\vec{r_1},z_1)\ ,
\label{135}
 \EE
%%%%%%%%%%%%%%%%%%%%%%%%%%%%%%%%%%%%%%%%%%%%%%%%%%%%%%%%%%%%%%%%%%%%%%%%
%
with the boundary condition
%
%%%%%%%%%%%%%%%%%%%%%%%%%%%%%%%%%%%%%%%%%%%%%%%%%%%%%%%%%%%%%%%%%%%%%%%%
 \BE
G_{\bar qq}(\vec{r_2},z_2;\vec{r_1},z_1)|_{z_2=z_1}=
\delta^2(\vec{r_1}-\vec{r_2})\, .
\label{136}
 \EE
%%%%%%%%%%%%%%%%%%%%%%%%%%%%%%%%%%%%%%%%%%%%%%%%%%%%%%%%%%%%%%%%%%%%%%%%
%
In Eq.~(\ref{135}) $\nu$ is the photon energy and the Laplacian
$\Delta_{r}$ acts on the coordinate $r$.

We start with the propagation of a $\bar qq$ pair in vacuum. The LC
potential $V_{\bar qq}(z_2,\vec{r_2},\alpha)$ in (\ref{135})
contains only the real part, which is responsible for the
interaction between the $q$ and $\bar{q}$. For the sake of simplicity we use
an oscillator form of this potential. Although more realistic models for the 
real part of the potential are available \cite{pirner-04,pirner-08}, however, 
solution of the corresponding Schr\"odinger equation 
for the light-cone Green function  is 
a challenge. Analytic solution has been known so far only for the oscillator 
potential. Otherwise one has to solve the Schr\"odinger equation numerically, 
which needs a dedicated study.

On the other hand, important is the mean $\bar qq$ transverse 
separation which is fitted to diffraction data. Any form of the potential must 
comply with this condition. The same restriction is imposed on the quark-gluon 
Fock states. The mean quark-gluon separation, which matters for shadowing, is 
fixed by high-mass diffraction data and should not be much affected by the 
choice of a model for the potential.

%%%%%%%%%%%%%%%%%%%%%%%%%%%%%%%%%%%%%

 \BE
{\rm Re}\,V_{\bar qq}(z_2,\vec{r_2},\alpha) =
\frac{a^4(\alpha)\,\vec{r_2}\,^2}
{2\,\nu\,\alpha(1-\alpha)}\ ,
\label{140}
 \EE
%%%%%%%%%%%%%%%%%%%%%%%%%%%%%%%%%%%%%%%%%%%%%%%
%
one can solve then two-dimensional Schr\" odinger equation
(\ref{135}) analytically, and the solution is given by the harmonic
oscillator Green function \cite{fg}
%
%%%%%%%%%%%%%%%%%%%%%%%%%%%%%%%%%%%%%%%%%%%%%%%%%%%%%%%%%%
 \BA
G_{\bar qq}(\vec{r_2},z_2;\vec{r_1},z_1) =
\frac{a^2(\alpha)}{2\;\pi\;i\;
{\rm sin}(\omega\,\Delta z)}\, {\rm exp}
\left\{\frac{i\,a^2(\alpha)}{{\rm sin}(\omega\,\Delta z)}\,
\Bigl[(r_1^2+r_2^2)\,{\rm cos}(\omega \;\Delta z) -
2\;\vec{r_1}\cdot\vec{r_2}\Bigr]\right\}
\nonumber\\ \times {\rm exp}\left[-
\frac{i\,\epsilon^{2}\,\Delta z}
{2\,\nu\,\alpha\,(1-\alpha)}\right] \ ,
\label{142}
 \EA
%%%%%%%%%%%%%%%%%%%%%%%%%%%%%%%%%%%%%%%%%%%%%%%%%%%%%%%%%%%%
%
where $\Delta z=z_2-z_1$, and
%
%%%%%%%%%%%%%%%%%%%%%%%%%%%%%%%%%%%%%%%%%%%%%%%%%%%%%%%%%%%%
 \BE \omega = \frac{a^2(\alpha)}{\nu\;\alpha(1-\alpha)}\ .
\label{144}
 \EE
%%%%%%%%%%%%%%%%%%%%%%%%%%%%%%%%%%%%%%%%%%%%%%%%%%%%%%%%%%%%
%
The shape of the function $a(\alpha)$ in Eq.~(\ref{140}) will be
discussed below.

The probability amplitude to find the $\bar qq$ fluctuation of a
photon at the point $z_2$ with separation $\vec r$, is given by an
integral over the point $z_1$ where the $\bar qq$ is created by the
photon with initial separation zero,
%
%%%%%%%%%%%%%%%%%%%%%%%%%%%%%%%%%%%%%%%%%%%%%%
 \BE
\Psi^{T,L}_{\bar qq}(\vec r,\alpha)=
\frac{i\,Z_q\sqrt{\alpha_{em}}}
{4\pi\,E\,\alpha(1-\alpha)}
\int\limits_{-\infty}^{z_2}dz_1\,
\Bigl(\bar\chi\;\widehat O^{T,L}\chi\Bigr)\,
G_{\bar qq}(\vec{r},z_2;\vec{r_1},z_1)
\Bigr|_{r_1=0}\ .
\label{146}
 \EE
%%%%%%%%%%%%%%%%%%%%%%%%%%%%%%%%%%%%%%%%%%%%%%
%
 The operators $\widehat O^{T,L}$ are defined by Eqs.~(\ref{124}) and
(\ref{125}), and here they act on the coordinate $\vec r_1$.

If we write the transverse part  as
%
%%%%%%%%%%%%%%%%%%%%%%%%%%%%%%%%%%%%%%%%%%%%%%%%%%%%%%%%%%%%%%%%%%
 \BE
\bar\chi\;\widehat O^{T}\chi=
\bar\chi\;m_{c}\,\,\vec{\sigma}\cdot\vec{e}\,\chi +
\bar\chi\;\left[i\,(1-2\alpha)\,(\vec{\sigma}\cdot\vec{n})\,
\vec{e} + (\vec{\sigma}\times
\vec{e})\right]\,\chi\cdot\vec{\nabla}_{r}=
E+\vec F\cdot\vec\nabla_{r}\ ,
\label{150}
 \EE
%%%%%%%%%%%%%%%%%%%%%%%%%%%%%%%%%%%%%%%%%%%%%%%%%%%%%%%%%%%%%%%%%%
%
 then the distribution functions read,
%
%%%%%%%%%%%%%%%%%%%%%%%%%%%%%%%%%%%%%%%%%%%%
 \BE
\Psi^{T}_{\bar qq}(\vec r,\alpha) =
Z_q\sqrt{\alpha_{em}}\,\left[E\,\Phi_0(\epsilon,r,\lambda)
+ \vec F\,\vec\Phi_1(\epsilon,r,\lambda)\right]\ ,
\label{152}
 \EE
%%%%%%%%%%%%%%%%%%%%%%%%%%%%%%%%%%%%%%%%%%%%
%
%
%%%%%%%%%%%%%%%%%%%%%%%%%%%%%%%%%%%%%%%%%%%%
 \BE
\Psi^{L}_{\bar qq}(\vec r,\alpha) =
2\,Z_q\sqrt{\alpha_{em}}\,Q\,\alpha(1-\alpha)\,
\bar\chi\;\vec\sigma\cdot\vec n\;\chi\,
\Phi_0(\epsilon,r,\lambda)\ ,
\label{154}
 \EE
%%%%%%%%%%%%%%%%%%%%%%%%%%%%%%%%%%%%%%%%%%%%
%
 where
%
%%%%%%%%%%%%%%%%%%%%%%%%%%%%%%%%%%%%%%
 \BE
\lambda=
\frac{2\,a^2(\alpha)}{\epsilon^2}\ .
\label{156}
 \EE
%%%%%%%%%%%%%%%%%%%%%%%%%%%%%%%%%%%%%%
%

The functions $\Phi_{0,1}$ in Eqs.~(\ref{152}) and (\ref{154})
are defined as
%
%%%%%%%%%%%%%%%%%%%%%%%%%%%%%%%%%%%%%%%%%%%%%%%%%%%
 \BE
\Phi_0(\epsilon,r,\lambda) =
\frac{1}{4\pi}\int\limits_{0}^{\infty}dt\,
\frac{\lambda}{{\rm sh}(\lambda t)}\,
{\rm exp}\left[-\ \frac{\lambda\epsilon^2 r^2}{4}\,
{\rm cth}(\lambda t) - t\right]\ ,
\label{160}
 \EE
%%%%%%%%%%%%%%%%%%%%%%%%%%%%%%%%%%%%%%%%%%%%%%%%%%%
%
%
%%%%%%%%%%%%%%%%%%%%%%%%%%%%%%%%%%%%%%%%%%%%%%%%%%%%%%%%%%%%%%%
 \BE
\vec\Phi_1(\epsilon,r,\lambda) =
\frac{\epsilon^2\vec r}{8\pi}\int\limits_{0}^{\infty}dt\,
\left[\frac{\lambda}{{\rm sh}(\lambda t)}\right]^2\,
{\rm exp}\left[-\ \frac{\lambda\epsilon^2 r^2}{4}\,
{\rm cth}(\lambda t) - t\right]\ ,
\label{162}
 \EE
%%%%%%%%%%%%%%%%%%%%%%%%%%%%%%%%%%%%%%%%%%%%%%%%%%%%%%%%%%%%%%%
%
where $sh(x)$ and $cth(x)$ are the hyperbolic sine and hyperbolic
cotangent, respectively.

Note that the $\bar q-q$ interaction enters in Eqs.~(\ref{152}) and
(\ref{154}) via the parameter $\lambda$ defined in Eq.~(\ref{156}).
In the limit of vanishing interaction $\lambda\to 0$ (i.e. $Q^2\to
\infty$, $\alpha$ is fixed, $\alpha\not=0$ or $1$) Eqs.~(\ref{152})
- (\ref{154})  produce the perturbative expressions of
Eq.~(\ref{122}).

With the choice $a^2(\alpha)\propto \alpha(1-\alpha)$, the end-point
behavior of the mean square interquark separation is $\la
r^2\ra\propto 1/\alpha(1-\alpha)$, which contradicts the idea of
confinement. Following \cite{kst2} we fix this problem via a simple
modification of the LC potential,
%
%%%%%%%%%%%%%%%%%%%%%%%%%%%%%%%%%%%%%%%%%%%%%%%%%%%
 \BE
a^2(\alpha) = a^2_0 +4a_1^2\,\alpha(1-\alpha)\ .
\label{180}
 \EE
%%%%%%%%%%%%%%%%%%%%%%%%%%%%%%%%%%%%%%%%%%%%%%%%%%%
%
 The parameters $a_0$ and $a_1$ were adjusted in \cite{kst2} to data on
total photoabsorption cross section \cite{gamma1,gamma2},
diffractive photon dissociation, and shadowing in nuclear
photoabsorption reaction. The results of our calculations vary
within only $1\%$ when $a_0$ and $a_1$ satisfy the relation,
%
%%%%%%%%%%%%%%%%%%%%%%%%%%%%%%%%%%%%%%%%%%%%%%%%%%%
 \BA
a_0^2&=&v^{ 1.15}\, (0.112)^2\,\GeV^{2}\nonumber\\
a_1^2&=&(1-v)^{1.15}\,(0.165)^2\,\GeV^{2}\ ,
\label{190}
 \EA
%%%%%%%%%%%%%%%%%%%%%%%%%%%%%%%%%%%%%%%%%%%%%%%%%%%
%
 where $v$ takes any value $0<v<1$. In view of this insensitivity of the
observables we fix the parameters at $v=1/2$. We checked that this
choice does not affect our results beyond a few percent uncertainty.

The matrix element (\ref{120}) contains the LC wave function
squared, which has the following form for T and L polarizations, in
the limit of vanishing interaction between $\bar q$ and $q$,
%
%%%%%%%%%%%%%%%%%%%%%%%%%%%%%%%%%%%%%%%%%%%%%%%%%%%%%%%%%%%
 \BE
\Bigl |\Psi^{T}_{\bar qq}(\vec r,\alpha,Q^2)\,\Bigr |^2 =
\frac{2\,N_C\,\alpha_{em}}{(2\pi)^2}\,
\sum_{f=1}^{N_f}\,Z_f^2
\left[m_f^2\,K_0(\epsilon,r)^2
+ [\alpha^2+(1-\alpha)^2]\,\epsilon^2\,K_1(\epsilon\,r)^2\right]\ ,
%----------
\label{197a}
%----------
 \EE
%%%%%%%%%%%%%%%%%%%%%%%%%%%%%%%%%%%%%%%%%%%%%%%%%%%%%%%%%%%
%
and
%
%%%%%%%%%%%%%%%%%%%%%%%%%%%%%%%%%%%%%%%%%%%%%%%%%%%%%%%%%%%
 \BE
\Bigl |\Psi^{L}_{\bar qq}(\vec r,\alpha,Q^2)\,\Bigr |^2 =
\frac{8\,N_C\,\alpha_{em}}{(2\pi)^2}\,
\sum_{f=1}^{N_f}\,Z_f^2
\,Q^2\,\alpha^2(1-\alpha)^2\,
K_0(\epsilon\,r)^2\ ,
%----------
\label{197b}
%----------
 \EE
%%%%%%%%%%%%%%%%%%%%%%%%%%%%%%%%%%%%%%%%%%%%%%%%%%%%%%%%%%%
%
where $K_1$ is the modified Bessel function,
%
%%%%%%%%%%%%%%%%%%%%%%%%%%%%%%%%%%%%%%%%%%%%%%%%%%%%%%%%%%%
 \BE
K_1(z) = - \frac{d}{dz}\,K_0(z)\, .
%----------
\label{198}
%----------
 \EE
%%%%%%%%%%%%%%%%%%%%%%%%%%%%%%%%%%%%%%%%%%%%%%%%%%%%%%%%%%%
%

If one includes the nonperturbative $\bar q-q$ interaction, the
perturbative expressions (\ref{197a}) and (\ref{197b}) should be
replaced by:
%
%%%%%%%%%%%%%%%%%%%%%%%%%%%%%%%%%%%%%%%%%%%%%%%%%%%%%%%%%%%
 \BE
\Bigl |\Psi^{T}_{npt}(\vec r,\alpha,Q^2)\,\Bigr |^2 =
{2\,N_C\,\alpha_{em}}\,
\sum_{f=1}^{N_f}\,Z_f^2
\left[m_f^2\,\Phi_0^2(\epsilon,r,\lambda)
+ [\alpha^2+(1-\alpha)^2]\,\bigl |\vec\Phi_1(\epsilon,r,\lambda)\,
\bigr |^2\,\right]\ ,
%----------
\label{199a}
%----------
 \EE
%%%%%%%%%%%%%%%%%%%%%%%%%%%%%%%%%%%%%%%%%%%%%%%%%%%%%%%%%%%
%
and
%
%%%%%%%%%%%%%%%%%%%%%%%%%%%%%%%%%%%%%%%%%%%%%%%%%%%%%%%%%%%
 \BE
\Bigl |\Psi^{L}_{npt}(\vec r,\alpha,Q^2)\,\Bigr |^2 =
{8\,N_C\,\alpha_{em}}\,
\sum_{f=1}^{N_f}\,Z_f^2
\,Q^2\,\alpha^2(1-\alpha)^2\,
\Phi_0^2(\epsilon,r,\lambda)\ .
%----------
\label{199b}
%----------
 \EE
%%%%%%%%%%%%%%%%%%%%%%%%%%%%%%%%%%%%%%%%%%%%%%%%%%%%%%%%%%%
%

Notice that in the LC formalism the photon wave function contains
also higher Fock states $|\bar qq\ra$, $|\bar qqG\ra$, $|\bar
qq2G\ra$, etc., but its effect can be implicitly incorporated into
the energy dependence of the dipole cross section $\sigma_{\bar
qq}(\vec{r},s)$, as is given in Eq.~(\ref{120}). The energy
dependence of the dipole cross section is naturally included in the
realistic KST parametrization Eq. (\ref{kst-1}).

Now we will continue with our discussion of DIS on nuclear targets,
and will study the propagation of a $\bar qq$ pair in nuclear
matter. Some work has already been done in this direction. In fact,
the derivation of the formula for nuclear shadowing, keeping only
the first shadowing term in Eq.~(\ref{110}),
$\Delta\sigma_{tot}(x_{Bj},Q^2) = \Delta\sigma_{tot}(\bar qq)$, can
be found in \cite{rtv-99}. This term represents the shadowing
correction for the lowest $\bar qq$ Fock state, and has the
following form
%
%%%%%%%%%%%%%%%%%%%%%%%%%%%%%%%%%%%%%%%%%%%%%%%%%%%%%%%%%%%
 \BE
\Delta\sigma_{tot}(x_{Bj},Q^2) = \frac{1}{2}~{Re}~\int d^2 b \int_{-\infty}^{
\infty} dz_1 ~\rho_{A}(b,z_1) \int_{z_1}^{\infty} dz_2~
\rho_A(b,z_2)
\int_{0}^{1} d\alpha ~A(z_1,z_2,\alpha)\, ,
%----------
\label{230}
%----------
 \EE
%%%%%%%%%%%%%%%%%%%%%%%%%%%%%%%%%%%%%%%%%%%%%%%%%%%%%%%%%%%
%
with
%
%%%%%%%%%%%%%%%%%%%%%%%%%%%%%%%%%%%%%%%%%%%%%%%%%%%%%%%%%%%
 \BE
A(z_1,z_2,\alpha)
= \int d^2 r_2 ~\Psi^{*}_{\bar qq}(\vec{r_2},\alpha,Q^2)
~\sigma_{\bar qq}(r_2,s) \int d^2 r_1
~G_{\bar qq}(\vec{r_2},z_2;\vec{r_1},z_1)
~\sigma_{\bar qq}(r_1,s)
~\Psi_{\bar qq}(\vec{r_1},\alpha,Q^2)\, .
%----------
\label{240}
%----------
 \EE
%%%%%%%%%%%%%%%%%%%%%%%%%%%%%%%%%%%%%%%%%%%%%%%%%%%%%%%%%%%
%
When nonpertubative interaction effects between the $\bar q$ and $q$
are explicitly included, one should replace in Eq.~(\ref{240})
$\Psi_{\bar qq}(\vec r,\alpha,Q^2)\Longrightarrow \Psi_{npt}(\vec
r,\alpha,Q^2)$ and $\Psi_{\bar qq}^*(\vec
r,\alpha,Q^2)\Longrightarrow \Psi_{npt}^*(\vec r,\alpha,Q^2)$.

In Eq.~(\ref{230}) $\rho_{A}({b},z)$ represents the nuclear density
function defined at the point with longitudinal coordinate $z$ and
impact parameter $\vec{b}$.
%
%****************************************************************
%************************ FIG.1 *********************************
%****************************************************************
 \begin{figure}[tbh]
\includegraphics{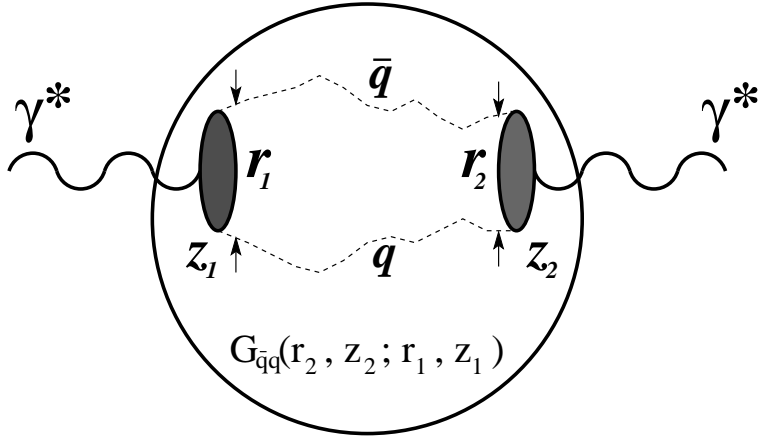}
\begin{center}
\vspace{6.7cm}
\parbox{13cm}
{\caption[Delta]
 {A cartoon for the shadowing term $\Delta\sigma_{tot}(x_{Bj},Q^2)
= \Delta\sigma_{tot}(\bar qq)$ in (\ref{110}). Propagation of the
$\bar qq$ pair through the nucleus is described by the Green
function $G_{\bar qq}(\vec{r_2},z_2;\vec{r_1},z_1)$, which results
from the summation over different paths of the $\bar qq$ pair.}
%%%%%%%%%%%%%%%%%%%%%%%%%
 \label{shad}}
%%%%%%%%%%%%%%%%%%%%%%%%%
\end{center}
 \end{figure}
%****************************************************************
%

The shadowing term $\Delta\sigma_{tot}(x_{Bj},Q^2) =
\Delta\sigma_{tot}(\bar qq)$ in (\ref{110}) is illustrated in
Fig.~\ref{shad}. At the point $z_1$ the initial photon diffractively
produces the $\bar qq$ pair ($\gamma^*N\to \bar qqN$) with
transverse separation $\vec{r_1}$. The $\bar qq$ pair then
propagates through the nucleus along arbitrary curved trajectories,
which are summed over, and arrives at the point $z_2$ with
transverse separation $\vec{r_2}$. The initial and final separations
are controlled by the LC wave function of the $\bar qq$ Fock
component of the photon $\Psi_{\bar qq}(\vec{r},\alpha,Q^2)$. During
propagation through the nucleus the $\bar qq$ pair interacts with
bound nucleons via the dipole cross section $\sigma_{\bar qq}(r,s)$,
which depends on the local transverse separation $\vec{r}$. The
Green function $G_{\bar qq}(\vec{r_2},z_2;\vec{r_1},z_1)$ describes
the propagation of the $\bar qq$ pair from $z_1$ to $z_2$.

Describing the propagation of the $\bar qq$ ~pair in a ~nuclear
~medium, the ~Green ~function ~$G_{\bar
qq}(\vec{r_2},z_2;\vec{r_1},z_1)$ satisfies again the time-dependent
two-dimensional Schr\"odinger equation (\ref{135}). However, the
potential in this case acquires in addition an imaginary part. This
imaginary part of the LC potential $V_{\bar qq}(z_2,\vec
r_2,\alpha)$ in Eq.~(\ref{135}) is responsible for the attenuation
of the $\bar qq$ photon fluctuation in the medium, and has the
following form
%
%%%%%%%%%%%%%%%%%%%%%%%%%%%%%%%%%%%%%%%%%%%%%%%%%%%%%%%%%%%%
 \BE
Im V_{\bar qq}(z_2,\vec r,\alpha) = -
\frac{\sigma_{\bar qq}(\vec r,s)}{2}\,\rho_{A}({b},z_2)\, .
%----------
\label{250}
%----------
 \EE
%%%%%%%%%%%%%%%%%%%%%%%%%%%%%%%%%%%%%%%%%%%%%%%%%%%%%%%%%%%%
%

As was already mentioned above, the analytical solution of
Eq.~(\ref{135}) is known only for the harmonic oscillator potential
$V_{\bar qq}(r)\propto r^2$. Consequently, in order to keep such an
analytical solution one should also use a quadratic approximation
for the imaginary part of $V_{\bar qq}(z_2,\vec r_2,\alpha)$, i.e.
%
%%%%%%%%%%%%%%%%%%%%%%%%%%%%%%%%%%%%%%%%%%
 \beq
\sigma_{\bar qq}(r,s) = C(s)\,r^2\ ,
%----------
\label{260}
%----------
 \eeq
%%%%%%%%%%%%%%%%%%%%%%%%%%%%%%%%%%%%%%%%%%
%
and uniform nuclear density
%
%%%%%%%%%%%%%%%%%%%%%%%%%%%%%%%%%%%%%%%%%%
 \beq
\rho_A(b,z) = \rho_0~\Theta(R_A^2-b^2-z^2)\, .
%----------
\label{270}
%----------
 \eeq
%%%%%%%%%%%%%%%%%%%%%%%%%%%%%%%%%%%%%%%%%%
%
In this case the solution of Eq.~(\ref{135}) has the same form as
Eq.~(\ref{142}), except that one should replace
$\omega\Longrightarrow \Omega$ and $a^2(\alpha)\Longrightarrow
b(\alpha)$, where
%
%%%%%%%%%%%%%%%%%%%%%%%%%%%%%%%%%%%%%%%%%%%%%
\BE
\Omega =
\frac{b(\alpha)}{\nu \alpha (1 - \alpha)}
=
\frac{
\sqrt{a^4(\alpha) - i\,\rho_A(b,z)\,\nu\,\alpha\,(1 - \alpha)\,C(s)}}
{\nu\,\alpha\,(1 - \alpha)}\, .
%----------
\label{280}
%----------
\EE
%%%%%%%%%%%%%%%%%%%%%%%%%%%%%%%%%%%%%%%%%%%%%
%

The determination of the energy dependent factor $C(s)$ in
Eq.~(\ref{260}) and the mean nuclear density $\rho_0$ in
Eq.~(\ref{270}) can be realized by the procedure described in
\cite{krt-00,r-00,n-03}, and will be discussed below.

Investigating nuclear shadowing in DIS one can distinguish between
two regimes, depending on the value of the coherence length:

{\bf (i)} We start with the general case when there are no
restrictions for $l_c$. If $l_c\sim R_A$ one has to take into
account the variation of the transverse size $r$ during propagation
of the $\bar qq$ pair through the nucleus, which is naturally
included using a correct quantum-mechanical treatment based on the
Green function formalism presented above. The overall total
photoabsorption cross section on a nucleus is given as a sum over T
and L polarizations, $\sigma^{\gamma^*A} = \sigma_T^{\gamma^*A} +
\epsilon'\,\sigma_L^{\gamma^*A}$, assuming that the photon
polarization $\epsilon'=1$. If one takes into account only the $\bar
qq$ Fock component of the photon, the full expression after
summation over all flavors, colors, helicities and spin states
becomes \cite{bgz-98}
%
%%%%%%%%%%%%%%%%%%%%%%%%%%%%%%%%%%%%%%%%%%%%%%%%%%%%%%%
\BA
\sigma^{\gamma^*A}(x_{Bj},Q^2) &=&
A\,\sigma^{\gamma^*N}(x_{Bj},Q^2) - \Delta\,\sigma(x_{Bj},Q^2)
\nonumber \\
&=& A\,\int\,d^2r\,\int_{0}^{1}\,d\alpha\,\sigma_{\bar qq}(r,s)
\,\Biggl (\Bigl |\Psi^T_{\bar qq}(\vec{r},\alpha,Q^2)\Bigr |^2 +
\Bigl |\Psi^L_{\bar qq}(\vec{r},\alpha,Q^2)\Bigr |^2\Biggr )
\nonumber \\
&-& \frac{N_C\,\alpha_{em}}{(2\pi)^2}\,\sum_{f=1}^{N_f}\,Z_f^2\,Re\,
\int\,d^2b\,\int_{-\infty}^{\infty}\,dz_1\,\int_{z_1}^{\infty}\,
dz_2\,\int_{0}^{1}\,d\alpha\,\int\,d^2r_1\,\int\,d^2r_2
\nonumber \\
&&\times\,\rho_A(b,z_1)\,\rho_A(b,z_2)\,\sigma_{\bar qq}(r_2,s)\,
\sigma_{\bar qq}(r_1,s)
\nonumber \\
&&\times\,\Biggl\{\Bigl[\,\alpha^2 + (1 - \alpha)^2\,\Bigr]
\,\epsilon^2
\,\frac{\vec{r_1}\,\cdot\,\vec{r_2}}{r_1\,r_2}\,
K_1(\epsilon\,r_1)\,K_1(\epsilon\,r_2)
%----------
\label{320}
%----------
\\
&&\,\,\,\,\,\,\,\,\,\,\,\,\,\,\,
 + \,\Bigl[\,m_f^2 + 4\,Q^2\,\alpha^2\,(1 - \alpha)^2\,\Bigr]\,
K_0(\epsilon\,r_1)\,K_0(\epsilon\,r_2)\Biggr\}\,
G_{\bar qq}(\vec{r_2},z_2;\vec{r_1},z_1) \, .
\nonumber
\EA
%%%%%%%%%%%%%%%%%%%%%%%%%%%%%%%%%%%%%%%%%%%%%%%%%%%%%%%
%
Here $\Bigl |\,\Psi^{T,L}_{\bar qq}(\vec{r},\alpha,Q^2)\,\Bigr |^2$
are the absolute squares of the LC wave functions for the $\bar qq$
fluctuation of T and L polarized photons, summed over all flavors,
and with the form given by Eqs.~(\ref{197a}) and (\ref{197b}),
respectively.

If one takes into account the nonperturbative interaction effects
between $\bar q$ and $q$ of the virtual photon the expression for
$\sigma^{\gamma^*A}(x_{Bj},Q^2)$ Eq.~(\ref{320}) takes the following
form:
%
%%%%%%%%%%%%%%%%%%%%%%%%%%%%%%%%%%%%%%%%%%%%%%%%%%%%%%%
\BA
\sigma^{\gamma^*A}_{npt}(x_{Bj},Q^2) &=&
A\,\sigma^{\gamma^*N}_{npt}(x_{Bj},Q^2) - \Delta\,\sigma_{npt}(x_{Bj},Q^2)
\nonumber \\
&=& A\,\int\,d^2r\,\int_{0}^{1}\,d\alpha\,\sigma_{\bar qq}(r,s)
\,\Biggl (\Bigl |\Psi^T_{npt}(\vec{r},\alpha,Q^2)\Bigr |^2 +
\Bigl |\Psi^L_{npt}(\vec{r},\alpha,Q^2)\Bigr |^2\Biggr )
\nonumber \\
&-& {N_C\,\alpha_{em}}\,\sum_{f=1}^{N_f}\,Z_f^2\,Re\,
\int\,d^2b\,\int_{-\infty}^{\infty}\,dz_1\,\int_{z_1}^{\infty}\,
dz_2\,\int_{0}^{1}\,d\alpha\,\int\,d^2r_1\,\int\,d^2r_2
\nonumber \\
&&\times\,\rho_A(b,z_1)\,\rho_A(b,z_2)\,\sigma_{\bar qq}(r_2,s)\,
\sigma_{\bar qq}(r_1,s)
\nonumber \\
&&
\times
\,\Biggl\{\Bigl[\,\alpha^2 + (1 - \alpha)^2\,\Bigr]
\,
\vec\Phi_1(\epsilon\,,r_1,\lambda)\cdot\vec\Phi_1(\epsilon\,,r_2,\lambda)
%----------
\label{325}
%----------
\\
&&\,\,\,\,\,\,\,\,\,\,\,\,\,
 + \,\Bigl[\,m_f^2 + 4\,Q^2\,\alpha^2\,(1 - \alpha)^2\,\Bigr]\,
\Phi_0(\epsilon\,,r_1,\lambda)\,\Phi_0(\epsilon\,,r_2,\lambda)\Biggr\}\,
G_{\bar qq}(\vec{r_2},z_2;\vec{r_1},z_1) \, .
\nonumber
\EA
%%%%%%%%%%%%%%%%%%%%%%%%%%%%%%%%%%%%%%%%%%%%%%%%%%%%%%%
%
where $\Bigl |\,\Psi^{T,L}_{npt}(\vec{r},\alpha,Q^2)\,\Bigr |^2$ are
now given by Eqs.~(\ref{199a}) and (\ref{199b}), respectively.

{\bf (ii)} The CL is much larger than the mean nucleon spacing in a
nucleus ($l_c\gg R_A$), which is the high energy limit.
Correspondingly, the transverse separation $r$ between $\bar q$ and
$q$ does not vary during propagation through the nucleus (Lorentz
time dilation). In this case the eikonal formula for the total
photoabsorption cross section on a nucleus can be obtained as a
limiting case of the Green function formalism. Indeed, in the high
energy limit $\nu\rightarrow \infty$, the kinetic term in
Eq.~(\ref{135}) can be neglected and the Green function reads
%
%%%%%%%%%%%%%%%%%%%%%%%%%%%%%%%%%%%%%%%%%%%%%%%%%%%%%
\BA
G_{\bar qq}(b;\vec{r_2},z_2;\vec{r_1},z_1)|_{\nu\to\infty} =
\delta(\vec{r_2}-\vec{r_1})\,\exp\Biggl[ - \frac{1}{2}\,
\sigma_{\bar qq}(r_2,s)\,\int_{z_1}^{z_2}\,dz\,\rho_A(b,z)\Biggr]\,.
%----------
\label{330}
%----------
\EA
%%%%%%%%%%%%%%%%%%%%%%%%%%%%%%%%%%%%%%%%%%%%%%%%%%%%%
%
Including nonperturbative interaction effects between $\bar q$ and
$q$, after substitution of the expression (\ref{330}) into
Eq.~(\ref{325}), one arrives at the following results:
%
%%%%%%%%%%%%%%%%%%%%%%%%%%%%%%%%%%%%%%%%%%%%%%%%%%%%%%%
\BA
\sigma^{\gamma^*A}_{npt}(x_{Bj},Q^2) &=&
2\,\int\,d^2b\,\int\,d^2r\,\int_0^1\,d\alpha
\left\{1 - \exp\,\Bigl[ - \frac{1}{2}\,\sigma_{\bar qq}(r,s)\,
T_A(b)\Bigr]\,\right\} \nonumber\\
&&\times\,{2\,N_C\,\alpha_{em}}\,\sum_{f=1}^{N_f}\,Z_f^2\,
\Biggl\{\Bigl[\,\alpha^2 + (1 - \alpha)^2\,\Bigr]
\,\Bigl |\vec\Phi_1(\epsilon\,,r,\lambda)\Bigr |^2\,
%----------
\label{335}
%----------
\\
&&\,\qquad\qquad\qquad\qquad
 + \,\Bigl[\,m_f^2 + 4\,Q^2\,\alpha^2\,(1 - \alpha)^2\,\Bigr]\,
\Phi_0^2(\epsilon\,,r,\lambda)\,\Biggr\}\, ,
\nonumber
\EA
%%%%%%%%%%%%%%%%%%%%%%%%%%%%%%%%%%%%%%%%%%%%%%%%%%%%%%%
%
where
%
%%%%%%%%%%%%%%%%%%%%%%%%%%%
\beq
T_A(b) = \int_{-\infty}^{\infty}\,dz\,\rho_A(b,z)\,
%-----------
\label{300}
%-----------
\eeq
%%%%%%%%%%%%%%%%%%%%%%%%%%%
%
is the nuclear thickness calculated with the realistic Wood-Saxon
form of the nuclear density, with parameters taken from
\cite{saxon}.

At the photon polarization parameter $\epsilon'=1$ the structure
function ratio 
%$F_2^A(x_{Bj},Q^2)/F_2^N(x_{Bj},Q^2)$ 
$F_2^A/F_2^N$ 
is related to nuclear shadowing $R(A/N$
and can be
expressed via a ratio of the total photoabsorption cross sections
%
%%%%%%%%%%%%%%%%%%%%%%%%%%%%%%
\BA
\frac{F_2^A(x_{Bj},Q^2)}{F_2^N(x_{Bj},Q^2)} =
A\,R(A/N)
=
\frac{\sigma_T^{\gamma^*A}(x_{Bj},Q^2)
+ \sigma_L^{\gamma^*A}(x_{Bj},Q^2)}
{\sigma_T^{\gamma^*N}(x_{Bj},Q^2)
+ \sigma_L^{\gamma^*N}(x_{Bj},Q^2)}\, ,
%----------
\label{340}
%----------
\EA
%%%%%%%%%%%%%%%%%%%%%%%%%%%%%%
%
where the numerator on the right-hand side (r.h.s.) is given by
Eq.~(\ref{325}), whereas the denominator can be expressed as the
first term of Eq.~(\ref{325}) divided by the mass number $A$.

As we already mentioned above, an explicit analytical expression for
the Green function $G_{\bar qq}(\vec{r_2},z_2;\vec{r_1},z_1)$
(\ref{142}) can be found only for the quadratic form of the dipole
cross section (\ref{260}), and for uniform nuclear density function
(\ref{270}). It was shown in refs.~\cite{krt-98,r-00,krt-00,n-03}
that such an approximation gives results of reasonable accuracy,
especially at small $x_{Bj}\lsim 10^{-4}$ and for heavy nuclei.
Nevertheless, it can be even more precise if one considers the fact
that the expression (\ref{335}) in the high energy limit can be
easily calculated using realistic parametrizations of the dipole
cross section (see Eq.~(\ref{kst-1}) for the KST parametrization and
ref.~\cite{gbw} for the GBW parametrization) and a realistic nuclear
density function $\rho_A(b,z)$ \cite{saxon}. Consequently, one needs
to know the full Green function only in the transition region from
non-shadowing ($x_{Bj}\sim 0.1$) to a fully developed shadowing
given when coherence length $l_c\gg R_A$, which corresponds to
$x_{Bj}\lsim 10^{-4}$ depending on the value of $Q^2$.
%Such an exact knowledge of the Green function
%is very important because this region is predominantly
%covered by available experimental data.
Therefore, the value of the energy dependent factor $C(s)$ in
Eq.~(\ref{260}) can be determined by the procedure described in
refs.~\cite{krt-00,r-00,knst-01}. According to this procedure, the
factor $C(s)$ is adjusted by demanding that calculations employing
the approximation (\ref{260}) reproduce correctly the results for
nuclear shadowing in DIS based on the realistic parametrizations of
the dipole cross section Eq.~(\ref{kst-1}) in the limit $l_c\gg
R_A$, when the Green function takes the simple form (\ref{330}).
Consequently, the factor $C(s)$ is fixed by the relation
%
%%%%%%%%%%%%%%%%%%%%%%%%%%%%%%%%%%%%%%%%%%%%%%%%%%%
\BA
\frac
{\int d^2\,b\,\int d^2\,r\,\Bigl
|\Psi_{\bar qq}(\vec{r},\alpha,Q^2)\Bigr |^2\,
\left\{1 - \exp\,\Bigl[ - \frac{1}{2}\,C(s)\,r^2\,
T_A(b)\Bigr]\,\right\}}
{\int d^2\,r\,
\Bigl |\Psi_{\bar qq}(\vec{r},\alpha,Q^2)\Bigr |^2\,
C(s)\,r^2}
\nonumber \\
=
\frac{\int d^2\,b\,\int d^2\,r\,\Bigl
|\Psi_{\bar qq}(\vec{r},\alpha,Q^2)\Bigr |^2\,
\left\{1 - \exp\,\Bigl[ - \frac{1}{2}\,\sigma_{\bar qq}(r,s)\,
T_A(b)\Bigr]\,\right\}}
{\int d^2\,r\,
\Bigl |\Psi_{\bar qq}(\vec{r},\alpha,Q^2)\Bigr |^2\,
\sigma_{\bar qq}(r,s)}\, .
%-----------
\label{224}
%-----------
\EA
%%%%%%%%%%%%%%%%%%%%%%%%%%%%%%%%%%%%%%%%%%%%%%%%%%%
%
Correspondingly, the value $\rho_0$ of the uniform nuclear density
(\ref{270}) is fixed in an analogous way using the following
relation
%
%%%%%%%%%%%%%%%%%%%%%%%%%%%
\beq
\int\,d^2\,b\,\Biggl [1 - exp\,\Biggl ( - \sigma_0\,\rho_0
\,\sqrt{R_A^2 - b^2}\,\Biggr )\,\Biggr ] =
\int\,d^2\,b\,\Biggl [1 - exp\,\Biggl ( -
\frac{1}{2}\,\sigma_0\,T_A(b)\,\Biggr )\,\Biggr ]\, ,
%-----------
\label{228}
%-----------
\eeq
%%%%%%%%%%%%%%%%%%%%%%%%%%%
%
where the value of $\rho_0$ was found to be practically independent
of the cross section $\sigma_0$, when this changed from 1 to 50$\mb$
\cite{krt-00,r-00}. Such a procedure for the determination of the
factors $C(s)$ and $\rho_0$ was applied also in
refs.~\cite{knst-01,n-02}, in the case of incoherent and coherent
production of vector mesons off nuclei.

In order to remove the above mentioned uncertainties the evolution
equation for the Green function was solved numerically for the first
time in ref.~\cite{n-03}. Such an exact solution can be performed
for arbitrary parametrization of the dipole cross section and for
realistic nuclear density functions, although the nice analytical
form for the Green function is lost in this case.

In the process of numerical solution of the Schr\"odinger equation
(\ref{135}) for the Green function $G_{\bar
qq}(\vec{r_2},z_2;\vec{r_1},z_1)$ with the initial condition
(\ref{136}), it is much more convenient to use the following
substitutions \cite{n-03}
%
%%%%%%%%%%%%%%%%%%%%%%%%%%%%%%%%%%%%%%%%%%%%%%%%
\BA
g_0(\vec{r_2},z_2;z_1,\lambda) =
\int\,d^2r_1\,\Phi_0(\epsilon\,,r_1,\lambda)\,\sigma_{\bar qq}(r_1,s)\,
G_{\bar qq}(\vec{r_2},z_2;\vec{r_1},z_1)\, ,
%----------
\label{350}
%----------
\EA
%%%%%%%%%%%%%%%%%%%%%%%%%%%%%%%%%%%%%%%%%%%%%%%%
%
and
%
%%%%%%%%%%%%%%%%%%%%%%%%%%%%%%%%%%%%%%%%%%%%%%%%
\BA
\frac{\vec{r_2}}{r_2}\,g_1(\vec{r_2},z_2;z_1,\lambda) =
\int\,d^2r_1\,\vec\Phi_1(\epsilon\,,r_1,\lambda)\,\sigma_{\bar qq}(r_1,s)\,
G_{\bar qq}(\vec{r_2},z_2;\vec{r_1},z_1)\, .
%----------
\label{360}
%----------
\EA
%%%%%%%%%%%%%%%%%%%%%%%%%%%%%%%%%%%%%%%%%%%%%%%%
%
After some algebra with Eq.~(\ref{135}) these new functions
~$g_0(\vec{r_2},z_2;z_1,\lambda)$ and
~$g_1(\vec{r_2},z_2;z_1,\lambda)$ can be shown to satisfy the
following evolution equations
%
%%%%%%%%%%%%%%%%%%%%%%%%%%%%%%%%%%%%%%%%%%%%%%%%%%%%%%%%%%%%%%%%%%%%%%
\BE
i\frac{d}{dz_2}\,g_0(\vec{r_2},z_2;z_1,\lambda)=
\left\{\frac{1}{2\,\mu_{\bar qq}}
\left[\epsilon^{2} -
\frac{\partial^2}{\partial\,r_2^2}
- \frac{1}{r_2}\,\frac{\partial}{\partial\,r_2}\right]
+ V_{\bar qq}(z_2,\vec r_2,\alpha)\right\}
g_0(\vec{r_2},z_2;z_1,\lambda)\
%----------
\label{370}
%----------
 \EE
%%%%%%%%%%%%%%%%%%%%%%%%%%%%%%%%%%%%%%%%%%%%%%%%%%%%%%%%%%%%%%%%%%%%%%
%
and
%
%%%%%%%%%%%%%%%%%%%%%%%%%%%%%%%%%%%%%%%%%%%%%%%%%%%%%%%%%%%%%%%%%%%%%%
\BE
i\frac{d}{dz_2}\,g_1(\vec{r_2},z_2;z_1,\lambda)=
\left\{\frac{1}{2\,\mu_{\bar qq}}
\left[\epsilon^{2} -
\frac{\partial^2}{\partial\,r_2^2}
- \frac{1}{r_2}\,\frac{\partial}{\partial\,r_2}
+ \frac{1}{r_2^2}\right]
+ V_{\bar qq}(z_2,\vec r_2,\alpha)\right\}
g_1(\vec{r_2},z_2;z_1,\lambda)\ ,
%----------
\label{380}
%----------
 \EE
%%%%%%%%%%%%%%%%%%%%%%%%%%%%%%%%%%%%%%%%%%%%%%%%%%%%%%%%%%%%%%%%%%%%%%
%
with the boundary conditions
%
%%%%%%%%%%%%%%%%%%%%%%%%%%%%%%%%
\beq
g_0(\vec{r_2},z_2;z_1,\lambda)|_{z_2=z_1}=
\Phi_0(\epsilon\,,r_2,\lambda)\,\sigma_{\bar qq}(r_2,s)
%----------
\label{390}
%----------
\eeq
%%%%%%%%%%%%%%%%%%%%%%%%%%%%%%%%
%
and
%
%%%%%%%%%%%%%%%%%%%%%%%%%%%%%%%%
\beq
g_1(\vec{r_2},z_2;z_1,\lambda)|_{z_2=z_1}=
\tilde\Phi_1(\epsilon\,,r_2,\lambda)\,\sigma_{\bar qq}(r_2,s)\, ,
%----------
\label{400}
%----------
\eeq
%%%%%%%%%%%%%%%%%%%%%%%%%%%%%%%%
%
where $\tilde\Phi_1(\epsilon\,,r,\lambda)$ is connected with
$\vec\Phi_1(\epsilon\,,r,\lambda)$ by the following relation:
%
%%%%%%%%%%%%%%%%%%%%%%%%%%%%%%%%
\beq
\vec\Phi_1(\epsilon\,,r,\lambda) =
\frac{\vec{r}}{r}\tilde\Phi_1(\epsilon\,,r,\lambda)\, .
%----------
\label{405}
%----------
\eeq
%%%%%%%%%%%%%%%%%%%%%%%%%%%%%%%%
%
In Eqs.~(\ref{370}) and (\ref{380}) the quantity
%
%%%%%%%%%%%%%%%%%%%
\beq
\mu_{\bar qq} = \nu\,\alpha\,(1-\alpha)
%----------
\label{410}
%----------
\eeq
%%%%%%%%%%%%%%%%%%%
%
plays the role of the reduced mass of the $\bar qq$ pair.

Now the expression (\ref{325}) for total photoabsorption cross
section on a nucleus reads
%
%%%%%%%%%%%%%%%%%%%%%%%%%%%%%%%%%%%%%%%%%%%%%%%%%%%%%%%
\BA
\sigma^{\gamma^*A}_{npt}(x_{Bj},Q^2) &=&
A\,\sigma^{\gamma^*N}_{npt}(x_{Bj},Q^2) - \Delta\,\sigma(x_{Bj},Q^2)
\nonumber \\
&=& A\,\int\,d^2r\,\int_{0}^{1}\,d\alpha\,\sigma_{\bar qq}(r,s)
\,\Biggl (\Bigl |\Psi^T_{npt}(\vec{r},\alpha,Q^2)\Bigr |^2 +
\Bigl |\Psi^L_{npt}(\vec{r},\alpha,Q^2)\Bigr |^2\Biggr )
\nonumber \\
&-& {3\,\alpha_{em}}\,\sum_{f=1}^{N_f}\,Z_f^2\,\,Re\,
\int\,d^2b\,\int_{-\infty}^{\infty}\,dz_1\,\int_{z_1}^{\infty}\,
dz_2\,\int_{0}^{1}\,d\alpha\,\int\,d^2r_2
\nonumber \\
&&\times\,\rho_A(b,z_1)\,\rho_A(b,z_2)\,\sigma_{\bar qq}(r_2,s)\,
\nonumber \\
&&\times\,\Biggl\{\Bigl[\,\alpha^2 + (1 - \alpha)^2\,\Bigr]\,
\tilde\Phi_1(\epsilon\,,r_2,\lambda)\,g_1(\vec{r_2},z_2;z_1,\lambda)
%----------
\label{420}
%----------
\\
&&\,\,\,\,\,\,\,\,\,\,\,\,\,\,\,
 + \,\Bigl[\,m_f^2 + 4\,Q^2\,\alpha^2\,(1 - \alpha)^2\,\Bigr]\,
\Phi_0(\epsilon\,,r_2,\lambda)\,g_0(\vec{r_2},z_2;z_1,\lambda)\Biggr\}\, .
\nonumber
\EA
%%%%%%%%%%%%%%%%%%%%%%%%%%%%%%%%%%%%%%%%%%%%%%%%%%%%%%%
%
Notice that this equation explicitly includes nonperturbative
interaction effects between $\bar q$ and $q$. Details of the
algorithm for the numerical solution of Eqs.~(\ref{370}) and
(\ref{380}) can be found in ref.~\cite{n-03}.

Finally we would like to emphasize that the $\bar qq$ Fock component
of the photon represents the highest twist shadowing correction
\cite{krt-00}, and vanishes at large quark masses as $1/m_f^2$. This
does not happen for higher Fock states containing gluons, which lead
to GS. Therefore GS represents the leading twist shadowing
correction \cite{kst2,kt-02}. Moreover, a steep energy dependence of
the dipole cross section $\sigma_{\bar qq}(r,s)$ (see
Eq.~(\ref{kst-1})) especially at smaller dipole sizes $r$ causes a
steep energy rise of both corrections.

%
%
%
%%%%%%%%%%%%%%%%%%%%%%%%%%%%%%%%%%%%%%%%%%%%%%%%%%%%%%%%%%
\section{Gluon shadowing}\label{glue-shadow}
%%%%%%%%%%%%%%%%%%%%%%%%%%%%%%%%%%%%%%%%%%%%%%%%%%%%%%%%%%
%
%
%

In the LC Green function approach
\cite{krt-98,krt-00,knst-01,n-02,n-03} the physical photon
$|\gamma^*\ra$ is decomposed into different Fock states, namely, the
bare photon $|\gamma^*\ra_0$, plus $|\bar qq\ra$, $|\bar qqG\ra$,
etc. As we mentioned above the higher Fock states containing gluons
describe the energy dependence of the photoabsorption cross section
on a nucleon, and also lead to GS in the nuclear case. However,
these fluctuations are heavier and have a shorter coherence time
(lifetime) than the lowest $|\bar qq\ra$ state, and therefore at
small and medium energies only the $|\bar qq\ra$ fluctuations of the
photon matter. Consequently, GS, which is related to the higher Fock
states, will dominate at higher energies, i.e. at small values of
$x_{Bj}\lsim 0.01$. Since we will study the shadowing region of
$x_{Bj}\lsim 0.01$ and the available experimental data reach values
of $x_{Bj}$ down to $\sim 10^{-4}$, we will include GS in our
calculations and show that it is not a negligible effect. Besides,
no data for gluon shadowing are available and one has to rely on
calculations.

In the previous Section~\ref{lc} we discussed the nuclear shadowing
for the $|\bar qq\ra$ Fock component of the photon. It is dominated
by the transverse photon polarizations, because the corresponding
photoabsorption cross section is scanned at larger dipole sizes than
for the longitudinal photon polarization. The transverse $\bar qq$
separation is controlled by the distribution amplitude
Eq.~(\ref{122}), with the mean value given by Eq.~(\ref{130}).
Contributions of large size dipoles come from the asymmetric $\bar
qq$ fluctuations of the virtual photon, when the quark and antiquark
in the photon carry a very large ($\alpha\to 1$) and a very small
fraction ($\alpha\to 0$) of the photon momentum, and vice versa. The
LC wave function for longitudinal photons (\ref{197b}) contains a
term $\alpha^2\,(1-\alpha)^2$, which makes considerably smaller the
contribution from asymmetric $\bar qq$ configurations than for
transversal photons (see Eq.~(\ref{197a})). Consequently, in
contrast to transverse photons, all $\bar qq$ dipoles from
longitudinal photons have a size squared $\propto 1/Q^2$ and the
double-scattering term vanishes as $\propto 1/Q^4$. The
leading-twist contribution for the shadowing of longitudinal photons
arises from the $|\bar qqG\ra$ Fock component of the photon because
the gluon can propagate relatively far from the $\bar qq$ pair,
although the $\bar q$-$q$ separation is of the order $1/Q^2$. After
radiation of the gluon the pair is in an octet state, and
consequently the $|\bar qqG\ra$ state represents a $GG$ dipole. Then
the corresponding correction to the longitudinal cross section is
just gluon shadowing.

The phenomenon of GS, just as for the case of nuclear shadowing
discussed in the Introduction, can be treated differently depending
on the reference frame. In the infinite momentum frame this
phenomenon looks similar to gluon-gluon fusion, corresponding to a
nonlinear term in the evolution equation \cite{glr}.  This effect
should lead to a suppression of the small-$x_{Bj}$ gluons also in a
nucleon, and lead to a precocious onset of the saturation effects
for heavy nuclei.  Within a parton model interpretation, in the
infinite momentum frame of the nucleus the gluon clouds of nucleons
which have the same impact parameter overlap at small $x_{Bj}$ in
the longitudinal direction. This allows gluons originated from
different nucleons to fuse, leading to a gluon density which is not
proportional to the density of nucleons any more. This is gluon
shadowing.

The same phenomenon looks quite different in the rest frame of the
nucleus. It corresponds to the process of gluon radiation and
shadowing corrections, related to multiple interactions of the
radiated gluons in the nuclear medium \cite{al}. This is a coherence
phenomenon known as the Landau-Pomeranchuk effect, namely the
suppression of bremsstrahlung by interference of radiation from
different scattering centers, demanding a sufficiently long
coherence time of radiation, a condition equivalent to a small
Bjorken $x_{Bj}$ in the parton model.

Although these two different interpretations are not Lorentz
invariant, they represent the same phenomenon, related to the
Lorentz invariant Reggeon graphs. It was already discussed in detail
in refs.~\cite{knst-01,krtj-02} that the double-scattering
correction to the cross section of gluon radiation can be expressed
in Regge theory via the triple-Pomeron diagram. It is interpreted as
a fusion of two Pomerons originated from different nucleons,
2$\,\Pom\rightarrow \Pom$, which leads to a reduction of the nuclear
gluon density $G_A$.

Notice that in the hadronic representation such a suppression of the
parton density corresponds to Gribov's inelastic shadowing
\cite{gribov}, which is related to the single diffraction cross
section. In particular, GS corresponds to the triple-Pomeron term in
the diffractive dissociation cross section, which enters the
calculations of inelastic corrections.

There are still very few numerical evaluations of gluon shadowing in
the literature, all of them done in the rest frame of the nucleus,
using the idea from ref.~\cite{al}. As was discussed above gluon
shadowing can be identified as the shadowing correction to the
longitudinal cross section coming from the $GG$ dipole representing
the $|\bar qqG\ra$ Fock component of the photon. An important point
for the evaluation of GS is knowing about the transverse size of
this $GG$ dipole. This size has been extracted in ref.~\cite{kst2}
from data for diffractive excitation of the incident hadrons to the
states of large mass, the so called triple-Pomeron region. The
corresponding diffraction cross section ($\propto r^4$) is a more
sensitive probe of the mean transverse separation than the total
cross section ($\propto r^2$). Consequently, it was found in
ref.~\cite{kst2} that the mean dipole size of the $GG$ system
(radius of propagation of the LC gluons) is rather small ,
$r_0\approx 0.3\,\fm$ \cite{k3p}. Such a small quark-gluon
fluctuation represents the only known way how to resolve the
long-standing problem of the small size of the triple-Pomeron
coupling.

To incorporate the smallness of the size of quark-gluon fluctuations
into the LC dipole approach, a nonperturbative LC potential
describing the quark-gluon interaction was introduced into the
Schr\"odinger equation for the LC Green function describing the
propagation of a quark-gluon system.  The strength of the potential
was fixed by data on high mass ($M_X^2$) diffraction $pp\to pX$
\cite{kst2}. This approach allows to extend the methods of pQCD to
the region of small $Q^2$. Since a new semihard scale $1/r_0 \sim
0.65\,\GeV$ is introduced, one should not expect a substantial
variation of gluon shadowing at $Q^2 \lsim 4/r_0^2$. Indeed, the
calculations performed in \cite{kst2} for $Q^2=0$ and $4\,\GeV^2$,
using different techniques, led to about the same gluon shadowing.
At higher $Q^2$ shadowing slowly (logarithmically) decreases, in
accordance with the expectations based on the evolution equation
\cite{mq-86}, which clearly demonstrates that GS is a leading-twist
effect.

In this paper we repeated the calculations \cite{kst2} of the ratio
of the gluon densities in nuclei and nucleon,
%
%%%%%%%%%%%%%%%%%%%%%%%%%%%%%%%%%%%%%%%%%%%%%%%%%%%%%%%%%%%%%%%%%%%%%%
 \beq
R_G(x_{Bj},Q^2)=\frac{G_A(x_{Bj},Q^2)}{A\,G_N(x_{Bj},Q^2)} \approx
1 - \frac{\Delta\sigma_{tot}(\bar qqG)}
{\sigma_{tot}^{\gamma^*A}}\ ,
\label{RG}
 \eeq
%%%%%%%%%%%%%%%%%%%%%%%%%%%%%%%%%%%%%%%%%%%%%%%%%%%%%%%%%%%%%%%%%%%%%%
%
where $\Delta\sigma_{tot}(\bar qqG)$ is the inelastic correction to
the total cross section $\sigma_{tot}^{\gamma^*A}$, related to the
creation of a $|\bar qqG\ra$ intermediate Fock state,
%
%%%%%%%%%%%%%%%%%%%%%%%%%%%%%%%%%%%%%%%%%%%%%%%%%%%%%%%%%%%%%
 \beqn
&& \Delta\sigma_{tot}(\bar qqG) =
{\rm Re}\int\limits_{-\infty}^{\infty}
dz_2 \int\limits_{-\infty}^{z_2} dz_1\,
\rho_A(b,z_1)\,\rho_A(b,z_2)
\int d^2x_2\,d^2y_2\,d^2x_1\,d^2y_1 \int
d\alpha_q\,\frac{d\,\alpha_G}{\alpha_G}
\nonumber\\ &\times&
F^{\dagger}_{\gamma^*\to\bar qqG}
(\vec x_2,\vec y_2,\alpha_q,\alpha_G)\
G_{\bar qqG}(\vec x_2,\vec y_2,z_2;\vec x_1,\vec y_1,z_1)\
F_{\gamma^*\to\bar qqG}
(\vec x_1,\vec y_1,\alpha_q,\alpha_G)\ .
\label{delta-sigma}
 \eeqn
%%%%%%%%%%%%%%%%%%%%%%%%%%%%%%%%%%%%%%%%%%%%%%%%%%%%%%%%%%%%%
%
 Here $\vec x$ and $\vec y$ are the transverse distances from the gluon
to the quark and antiquark, respectively, $\alpha_q$ is the fraction
of the LC momentum of the $\bar qq$ carried by the quark, and
$\alpha_G$ is the fraction of the photon momentum carried by the
gluon. $F_{\gamma^*\to\bar qqG}$ is the amplitude of diffractive
$\bar qqG$ production in a $\gamma^*N$ interaction \cite{kst2}, and
it is given by
%
%%%%%%%%%%%%%%%%%%%%%%%%%%%%%%%%%%%%%%%%%%%%%%%%%%%%%%%%%
 \beqn
F_{\gamma^*\to\bar qqG}(\vec x,\vec y,\alpha_q,\alpha_G)
&=& {9\over8}\,
\Psi_{\bar qq}(\alpha_q,\vec x -\vec y)\,
\left[\Psi_{qG}\left(\frac{\alpha_G}{\alpha_q},
\vec x\right) - \Psi_{\bar qG}
\left(\frac{\alpha_G}{1-\alpha_q},\vec y\right)\,
\right]\nonumber\\ &\times&
\biggl[\sigma_{\bar qq}(x)+
\sigma_{\bar qq}(y) -
\sigma_{\bar qq}(\vec x - \vec y)\biggr]\ ,
\label{amplitude}
\eeqn
%%%%%%%%%%%%%%%%%%%%%%%%%%%%%%%%%%%%%%%%%%%%%%%%%%%%%%%%%
%
 where $\Psi_{\bar qq}$ and $\Psi_{\bar qG}$ are the LC distribution
functions of the $\bar qq$ fluctuations of a photon and $qG$
fluctuations of a quark, respectively.

In the above equation $G_{\bar qqG}(\vec x_2,\vec y_2,z_2;\vec
x_1,\vec y_1,z_1)$ is the LC Green function which describes the
propagation of the $\bar qqG$ system from the initial state with
longitudinal and transverse coordinates $z_1$ and $\vec x_1,\vec
y_1$, respectively, to the final coordinates $(z_2,\vec x_2,\vec
y_2)$. For the calculation of gluon shadowing one should suppress
the intrinsic $\bar qq$ separation, i.e. assume $\vec x = \vec y$.
In this case the Green function simplifies, and effectively
describes the propagation of a gluon-gluon dipole through a medium.

An important finding of ref.~\cite{kst2} is the presence of a strong
nonperturbative interaction which squeezes the gluon-gluon wave
packet and substantially diminishes gluon shadowing. The smallness
of the gluon-gluon transverse separation is not a model assumption,
but is dictated by data for hadronic diffraction into large masses
(triple-Pomeron regime), which is controlled by diffractive gluon
radiation.

%
%****************************************************************
%************************ FIG.2 *********************************
%****************************************************************
 \begin{figure}[htb]
\includegraphics{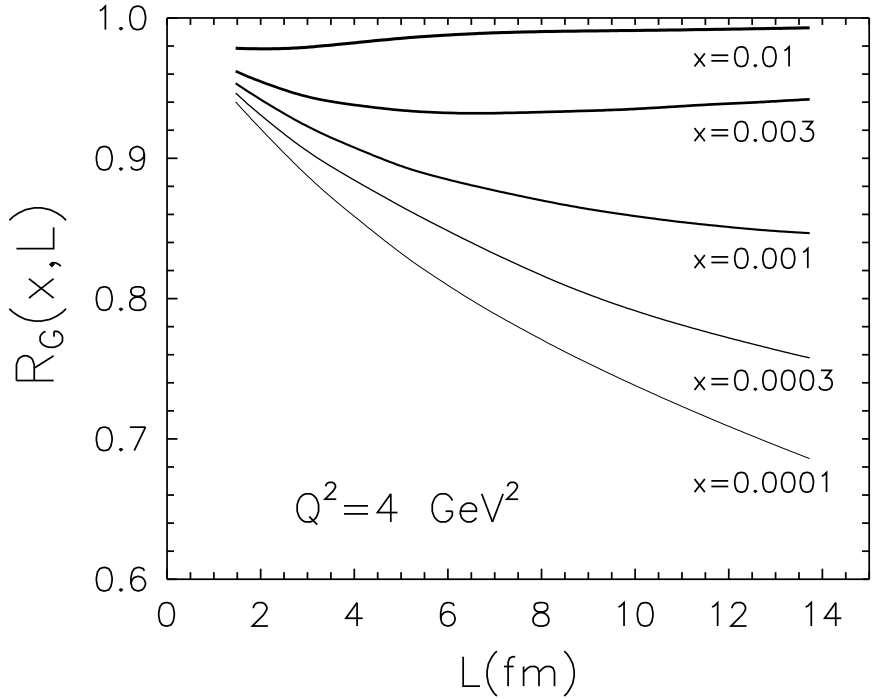}
\begin{center}
\vspace{9.2cm}
\parbox{14.0cm}
{\caption[Delta]
{
The ratio of the nucleus-to-nucleon gluon densities as function of the
thickness of the nucleus, $L=T(b)/\rho_0$, at $Q^2=4\,\GeV^2$ and 
different fixed
values of $x_{Bj}$. Figure is taken from ref.~\cite{knst-01}.
}
%%%%%%%%%%%%%%%%%%%%%%%%%
 \label{glue-shad}}
%%%%%%%%%%%%%%%%%%%%%%%%%
\end{center}
 \end{figure}
%****************************************************************
%

Further calculational details can be found in \cite{kst2}. In our
case we calculated the gluon shadowing only for the lowest Fock
component containing just one LC gluon. In terms of the parton model
it reproduces the effects of fusion of many gluons to one gluon (in
terms of Regge approach it corresponds to the $n\Pom \to \Pom$
vertex). Inclusion of higher multigluon Fock components is still a
challenge. However, their effect can be essentially taken into
account by the eikonalization of the calculated $R_G(x_{Bj},Q^2)$,
as argued in \cite{kth}. In other words, the dipole cross section,
which is proportional to the gluon density at small separations,
should be renormalized everywhere, in the form
%
%%%%%%%%%%%%%%%%%%%%%%%%%%%%%%%%%%%
 \beq
\sigma_{\bar qq} \Rightarrow
R_G\,\sigma_{\bar qq}\ .
\label{1100}
 \eeq
%%%%%%%%%%%%%%%%%%%%%%%%%%%%%%%%%%%
%
Such a procedure makes the nuclear medium more transparent. This
could be expected since Gribov's inelastic shadowing is known to suppress
the total hadron-nucleus cross sections, i.e. to make nuclei more
transparent \cite{zkl,kn}.

As an illustration of not very strong onset of GS, here we present
$R_G(x_{Bj},Q^2)$, Eq.~(\ref{RG}), for different nuclear thicknesses
$T_A(b)$.
Using an approximation of constant nuclear density (see
Eq.~(\ref{270})), $T_A(b)=\rho_0\,L$, where $L=2\,\sqrt{R_A^2 - b^2}$, the
ratio
$R_G(x_{Bj},Q^2)$ is also implicitly a function of $L$. An example for
the calculated $L$-dependence of $R_G(x_{Bj},Q^2)$ at $Q^2=4\,\GeV^2$ is
depicted in Fig.~\ref{glue-shad} for different values of $x_{Bj}$.

One can expect intuitively from Eq.~(\ref{1100}) that GS should
always diminish the nuclear cross sections of various processes in
nuclear targets, and that the onset of GS is stronger for heavier
nuclei. However, this is not so for incoherent electroproduction of
vector mesons, analyzed in ref.~\cite{knst-01}. The specific
structure of the expression for the nuclear production cross section
causes that the cross section of incoherent electroproduction of
vector mesons is rather insensitive to GS. Furthermore, the effect
of GS is stronger for light than for heavy nuclear targets, in
contradiction with the standard intuition. Moreover, for heavy
nuclei the effect GS can lead even to a counterintuitive enhancement
(antishadowing), as was analyzed in ref.~\cite{knst-01}. For the
case of coherent vector meson production $\gamma^*A \to VA$
\cite{knst-01}, GS was shown to be a much stronger effect in
comparison with incoherent production, which confirms the expected
reduction of the nuclear production cross section.

Similarly, it was analyzed in ref.~\cite{krtj-02} that multiple
scattering of higher Fock states containing gluons leads to an
additional suppression of the Drell-Yan cross section. In the
present paper we will demonstrate that gluon shadowing also
suppresses the total photoabsorption cross section on a nucleus
$\sigma_{tot}^{\gamma^* A}(x_{Bj},Q^2)$. Here we expect quite a
strong effect of GS in the shadowing region of small $x_{Bj}\lsim
(0.01\div 0.001)$, in the kinematic range of available data
corresponding to small and medium values of $Q^2\sim$ a few GeV$^2$.

%
%%%%%%%%%%%%%%%%%%%%%%%%%%%%%%%%%%%%%%%%%%%%%%%%%%%%%%%%%%
\section{Numerical results}
\label{results}
%%%%%%%%%%%%%%%%%%%%%%%%%%%%%%%%%%%%%%%%%%%%%%%%%%%%%%%%%%
%

As we mentioned above the main goal of this paper is to compare for
the first time available experimental data with realistic
predictions for nuclear shadowing in DIS, based on exact numerical
solutions of the evolution equation for the Green function. Such a
comparison is performed for the shadowing region of small
$x_{Bj}\lsim 0.01$. As was discussed in the previous section one
should take into account also a contribution of gluon shadowing,
which increases the overall nuclear suppression.  The effect of GS
was already calculated in ref.~\cite{j-02}, but only for the
$F_2^C/F_2^D$ ratio of structure functions. Although the quark
shadowing was compute approximately via longitudinal form factor of
the nucleus and assuming only the leading shadowing term, it was
shown that GS is a rather large effect at $x_{Bj}\sim 10^{-4}$. In
the present paper we will also show that GS is not a negligible
effect, and can in principle be detected by the data on the total
photoabsorption nuclear cross section in future experiments.

The predictions for nuclear shadowing based on an exact numerical
solution of the evolution equation was compared in ref.~\cite{n-03}
with approximate results obtained using the harmonic oscillatory
form of the Green function (\ref{142}). Quite a large discrepancy
was found, in the range of $x_{Bj}\gsim 0.001$, where the variation
of the transverse size of the $\bar qq$ pair during propagation
through the nucleus becomes important. Such a variation is naturally
included in the Green function formalism and consequently the exact
shape of the Green function is extremely important.

We use an algorithm for the numerical solution of the Schr\"odinger
equation for the Green function, as developed and described in
ref.~\cite{n-03}. This gives the possibility of calculating nuclear
shadowing for arbitrary LC potentials $V_{\bar
qq}(z,\vec{r},\alpha)$ and nuclear density functions. Because the
available data from the E665 \cite{e665-1,e665-2} and NMC
\cite{nmc-1,nmc-2} collaborations cover the region of small and
medium values of $Q^2 \lsim 4\,\GeV^2$, we prefer the KST
parametrization of the dipole cross section (\ref{kst-1}), which is
valid down to the limit of real photoproduction. On the contrary,
the second GBW parametrization \cite{gbw} of the dipole cross
section cannot be applied in the nonperturbative region and
therefore we do not use it in our calculations.

In the process of exact numerical solutions of the evolution
equation for the Green function, the imaginary part of the LC
potential (\ref{250}) contains the corresponding KST dipole cross
section as well. The nuclear density function $\rho_{A}({b},z)$ was
taken in the realistic Wood-Saxon form, with parameters taken from
ref.~\cite{saxon}. The nonperturbative interaction effects between
$\bar q$ and $q$ are included explicitly via the real part of the LC
potential of the form (\ref{140}), which is supported also by the
fact that the data from E665 and NMC collaborations correspond to
very small values of $Q^2 \lsim 1\,\GeV^2$ in the region of small
$x_{Bj}\lsim 0.004$.

%
%****************************************************************
%************************ FIG.3 *********************************
%****************************************************************
 \begin{figure}[tbh]
\includegraphics{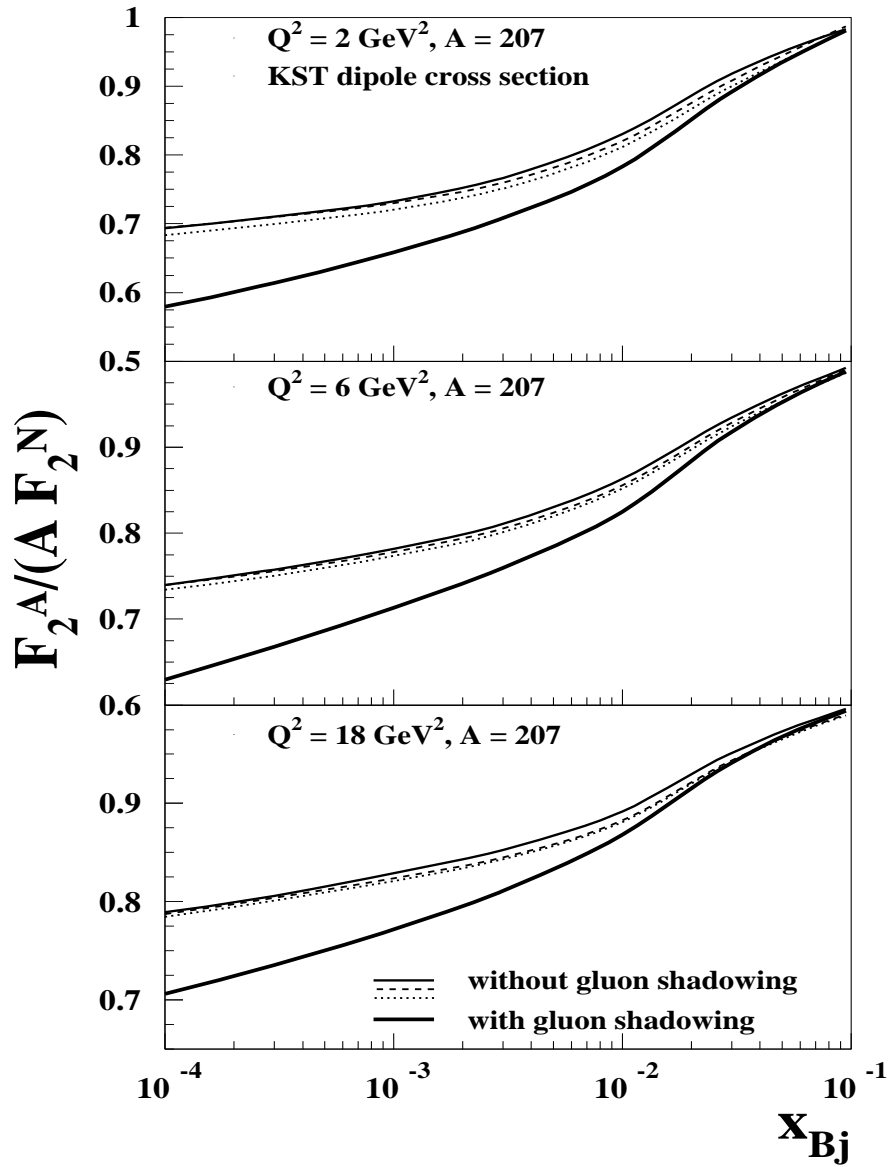}
\begin{center}
\vspace{15.5cm}
\parbox{13cm}
{\caption[Delta]
 {Nuclear shadowing for lead.
Calculations correspond to exact numerical solution of the evolution
equation for the Green function using the KST \cite{kst2}
parametrization of the dipole cross section and a realistic nuclear
density function of the Woods-Saxon form \cite{saxon}. The thick and
thin solid curves represent the predictions calculated with and
without contribution of gluon shadowing, respectively. The dotted
lines are calculated using a constant nuclear density function
(\ref{270}) and the quadratic form of the dipole cross section,
$\sigma(r,s) = C(s)\,r^2$, where the energy dependent factor C(s) is
determined by Eq.~(\ref{224}). The dashed curves are calculated for
the same quadratic form of the dipole cross section, but for the
realistic nuclear density function (\cite{saxon}) }
%%%%%%%%%%%%%%%%%%%%%%%%%
 \label{r-pb-all}}
%%%%%%%%%%%%%%%%%%%%%%%%%
\end{center}
 \end{figure}
%****************************************************************
%

We included also the effects of gluon shadowing for the lowest Fock
component containing just one LC gluon. Although the inclusion of
higher Fock components with more gluons is complicated, their effect
was essentially taken into account by eikonalization of the
calculated $R_G(x_{Bj},Q^2)$ \cite{kth}, i.e using the
renormalization (\ref{1100}).

Nuclear shadowing effects were studied via the $x_{Bj}$- behavior of
the ratio of proton structure functions (\ref{340}) divided by the
mass number $A$. First we present nuclear shadowing for a lead
target in Fig.~{\ref{r-pb-all}} at different fixed values of $Q^2$.
The thick and thin solid curves represent the predictions obtained
with and without the contribution of gluon shadowing, respectively.

%
%****************************************************************
%************************ FIG.4 *********************************
%****************************************************************
 \begin{figure}[tbh]
\includegraphics{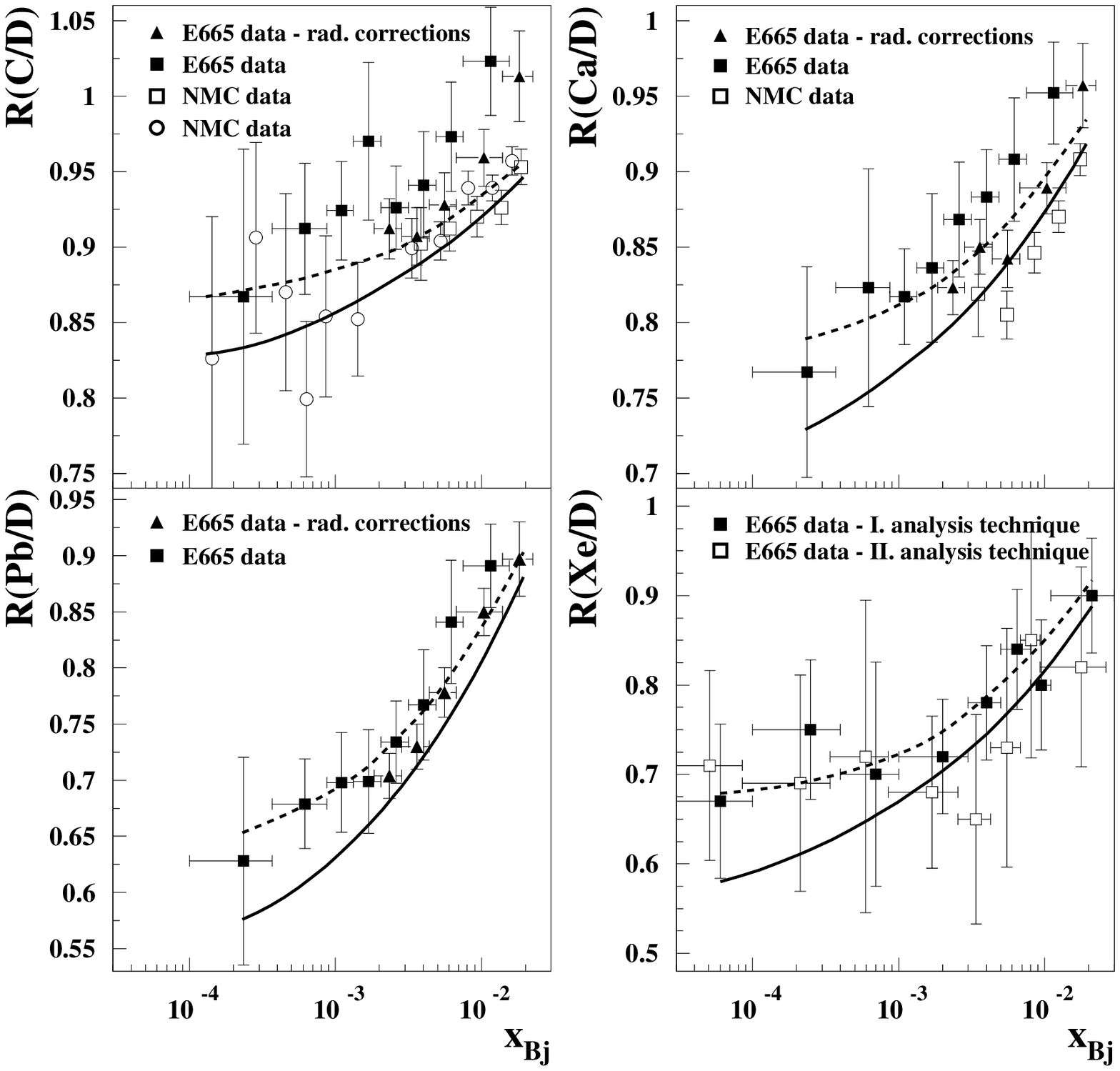}
\begin{center}
\vspace{12.0cm}
\parbox{13cm}
{\caption[Delta]
 {
Comparison of the model with experimental data from the E665
\cite{e665-1,e665-2} and NMC \cite {nmc-1,nmc-2} collaborations.
Calculations correspond to the exact numerical solution of the
evolution equation for the Green function using KST \cite{kst2}
parametrization of the dipole cross section and a realistic nuclear
density function of the Woods-Saxon form \cite{saxon}. The solid and
dashed curves are calculated with and without the contribution of
gluon shadowing, respectively. }
%%%%%%%%%%%%%%%%%%%%%%%%%
 \label{e665-all}}
%%%%%%%%%%%%%%%%%%%%%%%%%
\end{center}
 \end{figure}
%****************************************************************
%

One can see that the onset of GS happens at smaller $x_{Bj}$ than
the quark shadowing, which is supported by the fact that higher Fock
fluctuations containing gluons are in general heavier than $\bar qq$
and have a shorter coherence length. Fig.~\ref{r-pb-all}
demonstrates quite a strong effect of GS in the range of $x_{Bj}\in
(0.01,0.0001)$ where the most of available data exist. This is a
result of the suppression of the dipole cross section by the
renormalization (\ref{1100}), which can result only in a reduction
of the total photoabsorption cross section on a nuclear target.
Besides the effect of GS is stronger at smaller $Q^2$ because
corresponding Fock fluctuations of the photon have a larger
transverse size.

In Fig.~{\ref{r-pb-all}} we also present, for comparison and by the
dotted lines, the approximate predictions for nuclear shadowing in
DIS using constant nuclear density (\ref{270}) and the quadratic
form of the dipole cross section, $\sigma(r,s) = C(s)\,r^2$. The
energy dependent factor C(s) is determined by Eq.~(\ref{224}), and
the uniform nuclear density is fixed by the condition (\ref{228}).
One can see that these approximate predictions overestimate the
values of nuclear shadowing obtained by means of an exact numerical
solution of the evolution equation for the Green function. The
difference from the exact calculation (thin solid lines) is not
large and rises towards small values of $Q^2$. The reason is that
the quadratic approximation of the dipole cross section cannot be
applied exactly at large dipole sizes. Since the available data from
the E665 \cite{e665-1,e665-2} and NMC \cite{nmc-1,nmc-2}
collaborations at smallest values of $x_{Bj}$ correspond also to
small $Q^2\ll 1\,\GeV^2$, one can expect a larger difference between
the exact and approximate results in comparison with what is shown
in Fig.~{\ref{r-pb-all}} at $Q^2 = 2\,\GeV^2$. Keeping the quadratic
form of the dipole cross section, but using the realistic nuclear
density \cite{saxon}, one can obtain the results depicted in
Fig.~{\ref{r-pb-all} by the dashed lines. It brings a better
agreement with the exact calculations.

At low $x_{Bj}\lsim 10^{-4}$ one should expect a saturation of
nuclear shadowing at the level given by Eq.~(\ref{335}). This is
realized only for the dipole cross section, without energy
dependence, i.e. for example for parametrization (\ref{260}) of the
dipole cross section with constant factor $C(s)\approx 3$
\cite{krt-98}. However, this is not so for the realistic KST
parametrization Eq.~(\ref{kst-1}), where the saturation level is not
fixed exactly due to energy (Bjorken $x_{Bj}$-) dependence of the
dipole cross section $\sigma_{\bar qq}(r,s)$.

%
%****************************************************************
%************************ FIG.5 *********************************
%****************************************************************
 \begin{figure}[tbh]
\includegraphics{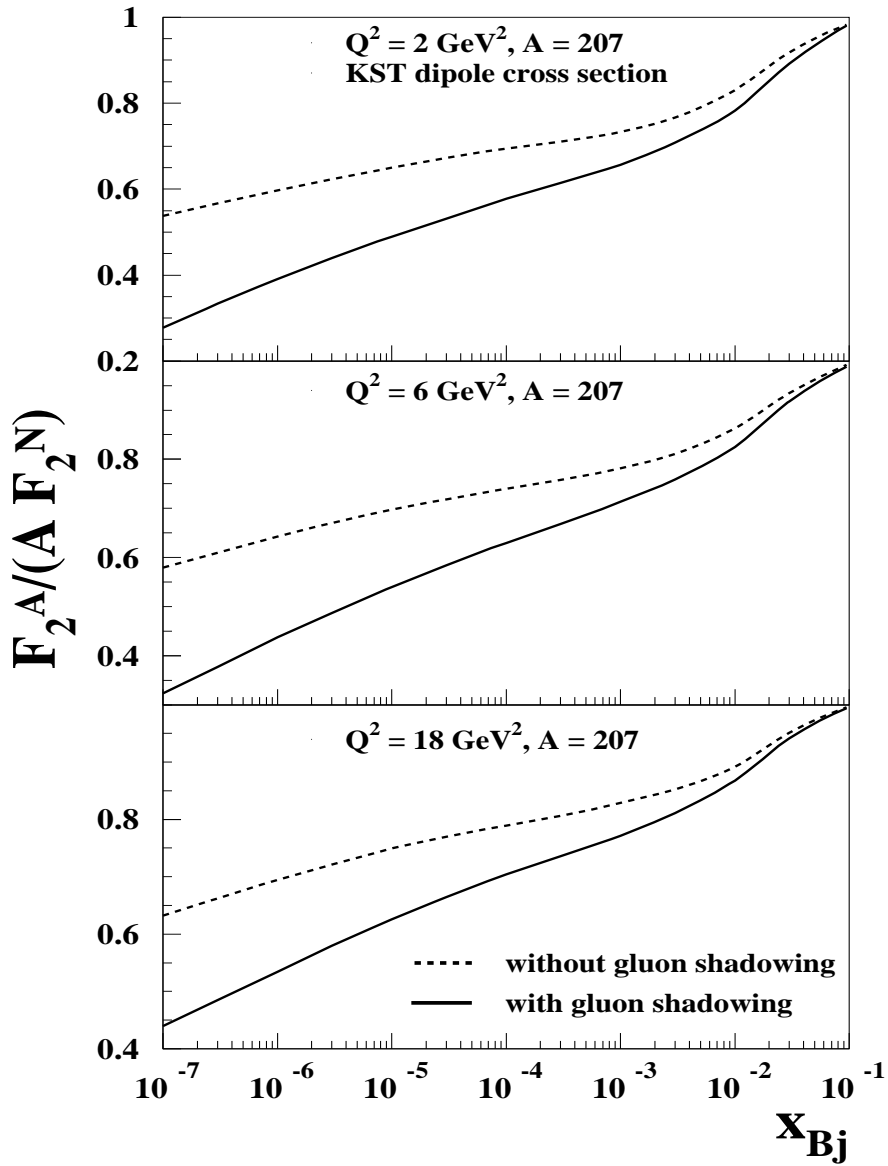}
\begin{center}
\vspace{15.0cm}
\parbox{16cm}
{\caption[Delta]
 {
Model predictions for nuclear shadowing 
for a broad $x_{Bj}$-region
down to $10^{-7}$ corresponding to 
LHC kinematical range at three different values
of $Q^2 = 2$, $6$ and $18\,\GeV^2$.
Calculations of the
nuclear shadowing for the $\bar qq$
Fock component of the photon 
correspond to the exact numerical solution of the
evolution equation for the Green function using KST \cite{kst2}
parametrization of the dipole cross section and a realistic nuclear
density function of the Woods-Saxon form \cite{saxon}. The solid and
dashed curves are calculated with and without the contribution of
gluon shadowing, respectively.
}
%%%%%%%%%%%%%%%%%%%%%%%%%
 \label{kst-LHC}}
%%%%%%%%%%%%%%%%%%%%%%%%%
\end{center}
 \end{figure}
%****************************************************************
%

In Fig.~\ref{e665-all} we present a comparison of the model
predictions with experimental data at small $x_{Bj}$, from the E665
\cite{e665-1,e665-2} and NMC \cite{nmc-1,nmc-2} collaborations. One
can see a quite reasonable agreement with experimental data, in
spite of the absence of any free parameters in the model. Several
comments are in order: first, if GS is not taken into account, for
the C/D and Ca/D ratios the nuclear shadowing looks overestimated in
comparison with the E665 data for $x_{Bj}\sim 0.01$, while it looks
in a good agreement for C/D, and a little bit underestimated for
Ca/D in comparison with the NMC data. This is affected by the known
incompatibility of the results from both experiments for the ratios
over D. Second, as was discussed in the previous section, the effect
of GS produces an additional nuclear shadowing which rises with mass
number $A$. Consequently, it leads to a small overestimation of the
nuclear shadowing for the C/D and Ca/D ratios in comparison with the
E665 data, but it seems to be a in good agreement with the NMC data.

%
%****************************************************************
%************************ FIG.6 *********************************
%****************************************************************
 \begin{figure}[tbh]
\includegraphics{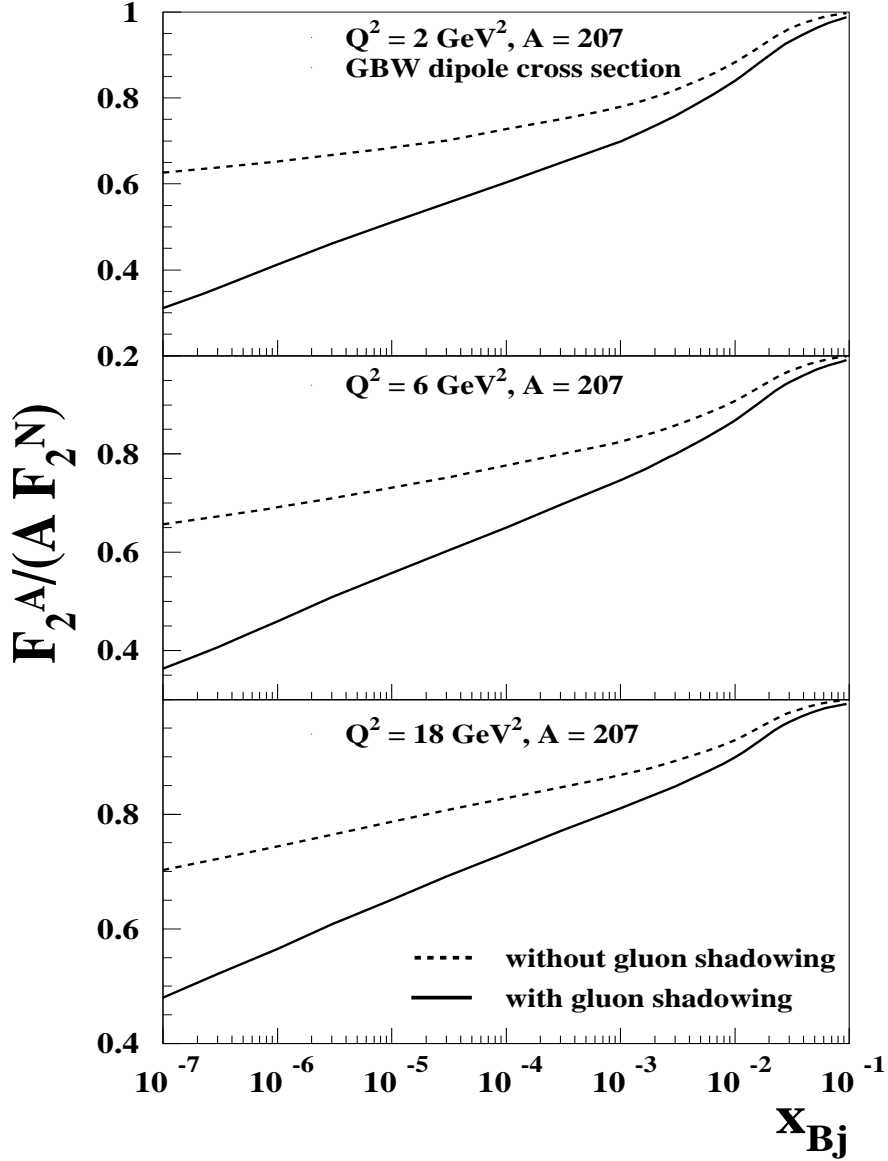}
\begin{center}
\vspace{15.0cm}
\parbox{16cm}
{\caption[Delta]
 {
The same as Fig.\ref{kst-LHC} but 
with GBW \cite{gbw}
parametrization of the dipole cross section.
}
%%%%%%%%%%%%%%%%%%%%%%%%%
 \label{gbw-LHC}}
%%%%%%%%%%%%%%%%%%%%%%%%%
\end{center}
 \end{figure}
%****************************************************************
%

For heavy nuclear targets there is only E665 data for the ratios
Xe/D and Pb/D. Fig.~\ref{e665-all} shows a reasonable good
description of these data, even if the effect of GS is taken into
account. The difference between the solid and dashed lines in
Fig.~\ref{e665-all} represents quite a large effect of GS, which was
neglected up to the present time in calculations of nuclear
shadowing in DIS \cite{krt-98,krt-00,r-00} assuming that it would be
a very small effect in the kinematic range covered by the available
experimental data. On the contrary, looking at Fig.~\ref{e665-all}
one can see that the effect of GS as an additional nuclear shadowing
cannot be neglected and should be included in calculations already
in the region of $x_{Bj}\lsim 0.01\div 0.001$. Very large error bars
especially at small $x_{Bj}\sim 10^{-4}$ do not allow to investigate
separately the effect of GS, and therefore more exact new data on
nuclear shadowing in DIS at small $x_{Bj}$ are very important for
further exploratory studies of the nuclear modification of structure
functions and also for gluon shadowing.

For completeness we present also
in Figs.~\ref{kst-LHC} and \ref{gbw-LHC}
predictions for
nuclear shadowing down to very small $x_{Bj} = 10^{-7}$
accesible by experiments at LHC using two different
realistic parametrizations of the dipole cross section,
KST \cite{kst2} and GBW \cite{gbw}. Again, one can see
quite large effect of GS as a difference between
the solid and dashed lines.

Here, we would like to emphasize that at $x_{Bj}\lsim 10^{-4}$
transverse size separations of the photon fluctuations
are ``frozen'' during propagation through nuclear medium and
one can use the simplified expressions, Eq.~(\ref{330}) and (\ref{335})
for calculation of nuclear shadowing.

%
%****************************************************************
%************************ FIG.7 *********************************
%****************************************************************
 \begin{figure}[tbh]
\includegraphics{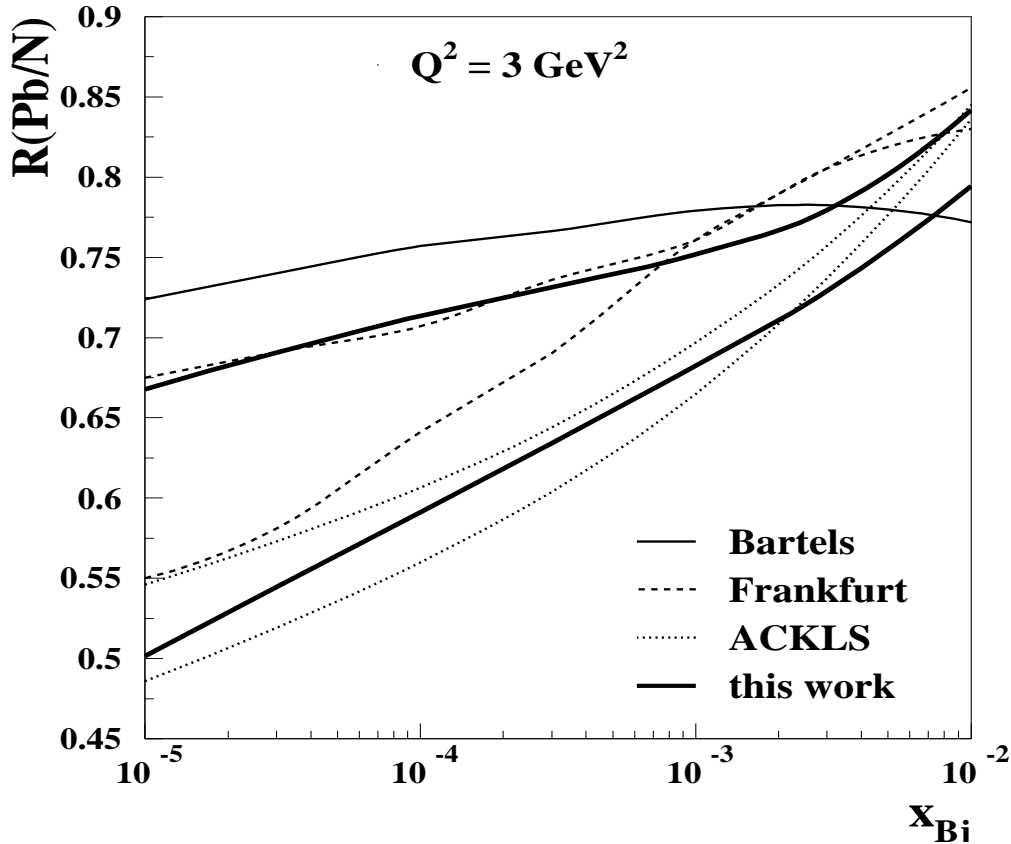}
\begin{center}
\vspace{12.0cm}
\parbox{13cm}
{\caption[Delta]
 {
Comparison of the model results for the ratio Pb/nucleon obtained
without (upper thick solid line) and with (lower thick solid line)
gluon shadowing,  with other models, versus $x_{Bj}$, at fixed $Q^2
= 3\,\GeV^2$. 
Bartels are the results from \cite{bartels}, 
Frankfurt from \cite{frankfurt} ($Q^2 = 4\,\GeV^2$), 
and ACKLS from \cite{ackls}. }
%%%%%%%%%%%%%%%%%%%%%%%%%
 \label{r-pb-3-all}}
%%%%%%%%%%%%%%%%%%%%%%%%%
\end{center}
 \end{figure}
%****************************************************************
%

Finally, we present in Fig.~\ref{r-pb-3-all} a comparison of the
nuclear shadowing calculated using our model with the results of
other models, for $Q^2 = 3\,\GeV^2$ (except the results of
ref.~\cite{frankfurt}, which are at $Q^2 = 4\,\GeV^2$). Notice that
the difference between models rises towards small values of
$x_{Bj}$, as a result of the different treatment of various nuclear
effects, and absence of relevant experimental information at such small $x_{Bj}$.
At $x_{Bj} = 10^{-5}$ we predict quite a large effect of GS (compare
upper and lower thick solid lines). 

In ref.~\cite{ackls} nuclear structure functions were studied using 
relation with diffraction on nucleons known as Gribov inelastic corrections.
The results of these calculations are depicted in Fig.~\ref{r-pb-3-all} by 
dotted curves.

The model presented in \cite{frankfurt} employs again
a parametrization of hard diffraction at the scale $Q_0^2$, which
gives nuclear shadowing in terms of Gribov's corrections similar
to ref.~\cite{ackls}. Then the nuclear suppression calculated at
$Q^2_0$ is used as initial condition for 
Dokshitzer-Gribov-Lipatov-Altareli-Parisi
(DGLAP)~\cite{dglap} evolution. This results are 
presented in Fig.~\ref{r-pb-3-all} by dashed curves.

The model based on a numerical solution of the non-linear equation for
small-$x_{Bj}$ evolution in nuclei was employed in \cite{bartels}. The result 
is shown in Fig.~\ref{r-pb-3-all} by thin solid curve.

At very small $x_{Bj}$ our model
predictions, including the effect of GS, roughly agrees with those of
\cite{ackls}, but lie below the results of other models. If the
effect of GS is not taken into account the situation is
substantially different and the corresponding curve (see upper thick
solid line) lies in between the results from other models.

Most models presented above are based on eikonal formulas, which
should be used only in the high energy limit, when the coherence
length $l_c \gg R_A$, i.e. at $x_{Bj}\lsim 10^{-4}$. However, they
were applied also in the region when $l_c \lsim R_A$. In this
transition shadowing region $x_{Bj}\in (0.0001,0.01)$ such
approximations lead in general to a larger nuclear shadowing than a
realistic situation, when more exact expressions should be more
appropriate (compare Eqs.~(\ref{325}) and (\ref{335})). For this
reason, theoretical predictions of most models overestimate nuclear
shadowing in the range of $x_{Bj}$ where available experimental data
exist.

%*********************************************************

So far the main source of experimental information on gluon shadowing was DIS 
on nuclei. Although it probes only quark distributions, the $Q^2$ dependence 
of nuclear effects is related via the evolution equations to the gluon distribution.
At the smallest value of $x_{Bj} = 0.01$ reached in the NMC experiment the 
gluon suppression factor $R_G(Sn)/R_G(C) = 0.87\pm 0.05$ was obtained in 
ref.~\cite{pirner-96} within the Leading-Log (LL) approximation. This result 
is somewhat lower that our expectation 
$R_G\sim 0.98$ which can be read out from Fig.~\ref{glue-shad}. However,
according to \cite{pirner-96} the next-to-LL corrections at 
$x_{Bj} = 0.01$ are about $10-20\%$, 
which apparently eliminates the disagreement with our calculations.
Furthermore, the full Leading-Order (LO) 
DGLAP analysis of the NMC data~\cite{nmc-1,nmc-2}
in ref.~\cite{eks98}, which should 
not be less accurate than  LL calculations, led to a conclusion that
the NMC data are not sensitive to gluon shadowing.
Moreover, the recent Next-to-LO (NLO) analysis by de Florian and Sassot
\cite{fs-nlo} was claimed to be sensitive to gluons. This analysis found 
almost no GS at $x_{Bj} = 0.01$ in good agreement with our calculations.

Other possible sources of information about gluon shadowing were considered in 
refs. \cite{arleo-1,arleo-2,eps-08}. It was proposed in \cite{arleo-1} to 
probe gluons in nuclei by direct photons produced in $p-A$ collisions in the 
proton fragmentation region where one can access smallest values of 
the light-front momentum fraction variable $x_2$ in nuclei.
This, however, should not work, since at large value of the light-front momentum 
fraction variable $x_1$ in the proton (i.e. at large Feynman $x_F$) 
one faces the energy sharing problem \cite{knpsj-05}: 
it is more difficult to give the whole energy to one particle in
$p-A$, than in $p-p$ collision. This effect leads to a breakdown of QCD 
factorization and to nuclear suppression 
observed at forward rapidities~\cite{knpsj-05,knpsj-05c,npps-08,trieste-08} 
in any reaction measured so far, even at low energies, where no shadowing is 
possible. 

An attempt to impose a restriction on GS
analyzing the nuclear effects in $J/\Psi$ production observed in $p-A$ 
collisions by the E866 experiment~\cite{e866-psi}, was made in \cite{arleo-2}. 
The mechanisms of $J/\Psi$ production and nuclear effects are so complicated,
that it would be risky to rely on oversimplified models. Indeed, the analysis
performed in this paper completely misses the color transparency 
effects, which are rather strong \cite{hhk} and vary throughout the interval 
of $x_F$ studied in this paper. For this reason the results of the analysis 
are not trustable. 

The new analysis of nuclear parton distribution functions performed in
~\cite{eps-08} included the  BRAHMS data for high-$p_T$ pion production at forward 
rapidities~\cite{brahms}. As we mentioned above, hadron production in this 
kinematic region of large $x_1$ ($x_F$) is suppressed by multiple parton 
interactions \cite{knpsj-05}, rather than by shadowing. Consequently, the 
results of this analysis are not trustable either. Moreover, it 
seems to provide another confirmation for an alternative dynamics for the 
suppression observed in the data \cite{brahms}. 
Indeed, it was concluded in \cite{eps-08} that gluons in lead target are 
completely terminated at $x_{Bj} = 10^{-4}$ where $R_G < 0.05$ is predicted. 
This is cannot be true because in the limit of strong shadowing the gluon ratio 
has a simple form $R_G = \pi R_A^2 / (A \sigma_{eff})$, where $\sigma_{eff}$ is 
the effective cross section responsible for shadowing. The strong effect 
predicted in \cite{eps-08} needs $\sigma_{eff}>150\mb$.

%***********************************************************
%
%%%%%%%%%%%%%%%%%%%%%%%%%%%%%%%%%%%%%%%%%%%%%%%%%%%%%%%%%%
\section{Summary and conclusions}
\label{conclusions}
%%%%%%%%%%%%%%%%%%%%%%%%%%%%%%%%%%%%%%%%%%%%%%%%%%%%%%%%%%
%

We presented a rigorous quantum-mechanical approach based on the
light-cone QCD Green function formalism which naturally incorporates
the interference effects of CT and CL. Within this approach
\cite{krt-98,rtv-99,krt-00,n-03} we studied nuclear shadowing in
deep-inelastic scattering at small Bjorken $x_{Bj}$.

Calculations of nuclear shadowing corresponding to the $\bar qq$
component of the virtual photon performed so far were based only on
efforts to solve the evolution equation for the Green function
analytically, and unfortunately an analytical harmonic oscillatory
form of the Green function (\ref{142}) could be obtained only by
using additional approximations, like a constant nuclear density
function (\ref{270}) and the dipole cross section of the quadratic
form (\ref{260}). This brings additional theoretical uncertainties
in the predictions for nuclear shadowing. In order to remove these
uncertainties we solve the evolution equation for the Green function
numerically, which does not require additional approximations.

In ref.~\cite{n-03} it was found for the first time the exact
numerical solution of the evolution equation for the Green function,
using two realistic parametrizations of the dipole cross section
(GBW \cite{gbw} and KST \cite{kst2}), and a realistic nuclear
density function of the Woods-Saxon form \cite{saxon}. It was
demonstrated that the corresponding nuclear shadowing shows quite
large differences from approximate results \cite{krt-98,krt-00}. On
the other hand, we showed that approximate calculations
corresponding to uniform nuclear density (\ref{370}) and quadratic
dipole cross section (\ref{360}), but with energy dependent factor
$C(s)$ determined by Eq.~(\ref{224}), bring a better agreement with
exact realistic calculations (see Fig.~\ref{r-pb-all}). However, the
difference from the exact calculations rises towards small values of
$Q^2$, where available data exist at smallest values of $x_{Bj}\sim
10^{-4}$. This confirms the claim that the quadratic approximation
of the dipole cross section cannot be applied at large dipole sizes.

Since the available data from the shadowing region of $x_{Bj}\lsim
0.01$ comes mostly from the E665 and NMC collaborations, and cover
only small and medium values of $Q^2\lsim 4\,\GeV^2$, we used only
the KST realistic parametrization \cite{kst2} of the dipole cross
section, which is more suitable for this kinematic region and the
corresponding expressions can be applied down to the limit of real
photoproduction. On the other hand, the data obtained at the lower
part of the $x_{Bj}$-kinematic interval correspond to very low
values of $Q^2 < 1\,\GeV^2$ (nonperturbative region). For this
reason we include explicitly the nonperturbative interaction effects
between $\bar q$ and $q$, taking into account the real part of the
LC potential $V_{\bar qq}$ (\ref{140}) in the time-dependent
two-dimensional Schr\"odinger equation (\ref{135}).

In order to compare the realistic calculations with data on nuclear
shadowing, the effects of GS are taken into account. The same path
integral technique \cite{kst2} can be applied in this case, and GS
was calculated only for the lowest Fock component containing just
one LC gluon. Although the inclusion of higher Fock components
containing more gluons is still a challenge, their effect was
essentially taken into account by eikonalization of the calculated
$R_G(x_{Bj},Q^2)$ ,using the renormalization (\ref{1100}). We found
quite a large effect of GS, which starts to be important already at
$x_{Bj}\sim 0.01$. The effect of GS rises towards small $x_{Bj}$
because higher Fock components with more gluons having shorter
coherence time will contribute to overall nuclear shadowing. Such a
situation is illustrated in Fig.~\ref{r-pb-all}.

Performing numerical calculations, we find that our model is in
reasonable agreement with existing experimental data (see
Fig.~\ref{e665-all}). Large error bars and incompatibility of the
experimental results from the E665 and NMC collaborations do not
allow to study separately the effect of GS, and therefore more
accurate new data on nuclear shadowing in DIS off nuclei at still
smaller $x_{Bj}\lsim 10^{-5}$ are very important for further
exploratory studies of GS effects.

Comparison among various models shows large differences for the
Pb/nucleon ratio of structure functions at $x_{Bj} = 10^{-5}$ and
$Q^2 = 3\,\GeV^2$ (see Fig.~\ref{r-pb-3-all}), which has a large
impact on the calculation of high-$p_T$ particles in  nuclear
collisions at RHIC and LHC. Such large differences at small $x_{Bj}$
among different models should be testable by the new more precise
data on nuclear structure functions, which can be obtained in
lepton-ion collider planned at BNL \cite{eRHIC}.

In most models presented above the final formulae for nuclear
shadowing are based on the eikonal approximation, which can be used
exactly only in the high energy limit, $l_c \gg R_A$. Consequently,
such an approach cannot be really applied in the transition
shadowing region, $l_c \approx R_A$, where $x_{Bj}\in
(0.0001,0.01)$, because it produces a larger nuclear shadowing than
in a realistic case when more appropriate expressions should be
taken into account (compare Eqs.~(\ref{325}) and (\ref{335})).

Concluding, a combination of the exact numerical solution of the
evolution equation for the Green function with the universality of
the LC dipole approach based on the Green function formalism
provides us with a very powerful tool for realistic calculations of
many processes. It allows to minimize theoretical uncertainties in
the predictions of nuclear shadowing in DIS off nuclei, which gives
the possibility to obtain reliable information about nuclear
modification of the structure functions at low $x_{Bj}$, with an
important impact on the physics performed in heavy ion collisions at
RHIC and in lepton-ion interactions planned at BNL.

\bigskip

{\bf{Acknowledgments:}}
This work was supported in part by Fondecyt (Chile) grants 1050589 and 1050519,
by DFG (Germany)  grant PI182/3-1, by the Slovak Funding
Agency, Grant No. 2/7058/27 and the grant VZ MSM 6840770039 and LC 07048 
(Czech Republic).

%%%%%%%%%%%%%%%%%%%%%%%%%%%%%%%%%%%%%%%%%%%%%%%%%%%%%%%%%%
 \def\appendix{\par
 \setcounter{section}{0} \setcounter{subsection}{0}
 \def\thesection{Appendix \Alph{section}}
\def\thesubsection{\Alph{section}.\arabic{subsection}}
\def\theequation{\Alph{section}.\arabic{equation}}
\setcounter{equation}{0}}
%%%%%%%%%%%%%%%%%%%%%%%%%%%%%%%%%%%%%%%%%%%%%%%%%%%%%%%%%%

\end{document}